\documentclass[12pt]{elsarticle}

\usepackage{graphics}
\usepackage{graphicx}

\usepackage{amssymb}

\usepackage{algorithm}
\usepackage{algorithmic}
\usepackage{url}
\usepackage{booktabs}
\usepackage{wrapfig}
\usepackage{listings}
\usepackage{amsthm}
\usepackage{array}
\usepackage{pbox}
\graphicspath{{images/}}

\journal{Information Systems}

\begin{document}

\begin{frontmatter}



\title{Cost optimization of data flows \\based on task re-ordering}


\author{Georgia Kougka}
\ead{georkoug@csd.auth.gr}
\address{Aristotle University of Thessaloniki}

\author{Anastasios Gounaris}
\ead{gounaria@csd.auth.gr}
\address{Aristotle University of Thessaloniki}

\begin{abstract}
Analyzing big data in a highly dynamic environment becomes more and more critical because of the increasingly need for end-to-end processing of this data. Modern data flows are quite complex and there are not efficient, cost-based, fully-automated, scalable optimization solutions that can facilitate flow designers. The state-of-the-art proposals fail to provide near optimal solutions even for simple data flows. To tackle this problem, we introduce a set of approximate algorithms for defining the execution order of the constituent tasks, in order to minimize the total execution cost of a data flow. We also present the advantages of the parallel execution of data flows. We validated our proposals in both a real tool and synthetic flows and the results show that we can achieve significant speed-ups, moving much closer to optimal solutions.
\end{abstract}


\end{frontmatter}


\section{Introduction}
\label{intro}
Data analysis in a highly dynamic environment becomes more and more critical in order to extract high-quality information from raw data that is nowadays produced at an extreme scale. The ultimate goal is to derive actionable information in a timely manner. To this end, we typically employ fully automated data-centric flows (or simply called data flows) both for business intelligence  \cite{CDN11} and scientific purposes \cite{ODOPVM11}, which typically execute under demanding performance requirements, e.g., to complete in a few seconds. Meeting such requirements, combined with the volatile nature of the environment and the data, gives rise to the need for efficient optimization techniques tailored to data flows.

Data flows define the processing of large data volumes as a sequence of data manipulation tasks. An example of a real-world, analytic flow is one that processes free-form text data retrieved from Twitter (tweets) that comment on products in order to compose a dynamic report considering sales, advertisement campaigns and user feedback after performing a dozen of steps \cite{Sim12}. Example steps include extraction of date information, quantifying the user sentiment through text analysis, filtering, grouping and expanding the information contained in the tweets through lookups in (static) data sources. Another example is to process newspaper articles, perform linguistic analysis, extract named entities and then establish relationships between companies and persons \cite{RHHLN13}. The tasks in a flow can either have a direct correspondence to relational operators, such as filters, grouping, aggregates and joins, or encapsulate arbitrary data transformations, text analytics, machine learning algorithms and so on
\cite{OCS09,HPS+12,Sim12}.

One of the most important steps in the data flow design is the specification of the execution order of the constituent tasks. In practice, this is usually the result of a manual procedure, which, in many cases results in non-optimal flow execution plans. Furthermore, even if a data flow is optimal for a specific input data set, it may prove significantly suboptimal for another data set with different characteristics \cite{HDP14}. We tackle this problem through the proposal of optimization algorithms that can provide the optimal execution order of the tasks in a data flow in an efficient manner and relieve the flow designers from the burden of selecting the task ordering on their own.  We consider a single optimization objective, namely the minimization of the sum of the task execution costs; we assume that the execution cost of each task depends on the volume of data to be processed, which in turn depends on the relative position of the task in the execution flow. The main challenges in flow optimization
that need to be addressed and differentiate the problem from that of traditional query optimization are as follows:
\begin{enumerate}
\item No arbitrary task orderings are valid, which means that the optimization algorithms need to respect the precedence constraints among tasks. E.g., in the introductory example, we cannot move a task that computes the average sentiment value from tweets before executing the task that quantifies the sentiment of the user through text analysis.
\item Flows can be very large with many constituent tasks, e.g., up to one hundred.
\end{enumerate}

The main implication is that query optimization techniques, which operate on plans with up to a few tens of operators that belong to the relational algebra (according to which operator reordering is typically permitted), are not applicable \cite{Cha98,Ioa96}. Nevertheless, they are  successful in their domain and this is the reason the data flow solutions proposed in this work are partially inspired by query optimization as we explain later. Overall, to date, there are very few proposals that deal with (or are applicable to) task reordering in data flows \cite{671SVS05,YLUG99,HPS+12}. A common characteristic of these proposals is that they are too slow to find an exact solution in small flows \cite{HPS+12}, or they can find significantly suboptimal (approximate) solutions for bigger flows \cite{671SVS05,YLUG99}.

In this work, we go beyond the state-of-the-art; we present both approximate and exact solutions. The approximate solutions are applicable to large flows and attain significantly better performance (more than 2 times speed-up in some settings, whereas in stand-alone cases, the speed-up is two or three orders of magnitude). The exact solution that we propose, although it cannot scale in general, it can process larger flows than those currently amenable to exact optimization. Our solutions apply to flows comprising any type of tasks and require as input common metadata that is task-independent, such as the average task selectivity and the task cost per invocation (e.g., in time units). Initially, we target linear flows, that is flows that can be described as a chain of activities with a single source and a single sink task; later, we relax this assumption.
The proposed optimization solutions were validated, as a proof of concept, in a real environment, namely Pentaho Data Integration (\emph{PDI}), which is a widespread data flow tool \cite{Pentaho}. Additionally, we performed thorough evaluations against existing approaches for synthetic data flows. The  summary of our contributions is as follows:
\begin{enumerate}
\item We provide a case study of data flow optimization implemented in PDI to provide insights into the inefficiency of the existing approaches and the actual benefits of our approaches (Section \ref{pentaho}).

\item We show that, under certain conditions, it is practical to derive optimal linear flows, even when the number of tasks is relatively large. Contrary to the case of query optimization, the most efficient solutions are those that leverage algorithms  enumerating valid topological orderings rather than dynamic programming or backtracking techniques (Section \ref{accurate}).

\item We introduce novel approximate low complexity algorithms that can be used for task reordering in data flows that have the form of a chain (Section \ref{approximate}).

\item  We discuss algorithms that produce flow execution plans, where a task sends its output to several downstream tasks in parallel; such an approach is suitable when the task selectivities are above 1, and can further improve on the performance of the flow execution plans (Section \ref{sec:parallel}).

\item We show how we can extend the solutions mentioned above to non linear flows with arbitrary number of sources and sinks (Section \ref{sec:mimo}).

\item We conduct thorough experiments in synthetic flows to detect the best optimization algorithm for linear and non-linear data flows among all of our proposals (Section \ref{experiments}). The evaluation results prove that the approaches introduced here significantly and consistently outperform the state-of-the-art in all out experiments.
\end{enumerate}

An extended abstract of some of the ideas above appears in \cite{KG14}.

%

\section{Problem Statement and Background}
\label{sec:background}

In this paper, we deal with the problem of re-ordering the tasks of a data flow without violating possible precedence constraints between tasks, while the performance of the flow is maximized. The data flow is represented as a directed acyclic graph (DAG), where each task corresponds to a node in the graph and the edges between nodes represent intermediate data shipping among tasks; i.e., in data flows, the exchange of data between tasks is explicitly represented through edges. The main notation, terminology and assumptions are as follows:
\renewcommand{\labelitemi}{$\bullet$}
\begin{itemize}
\item Let $G=(T,E)$ be a directed acyclic graph, where $T$ denotes the nodes of the graph (that correspond to flow tasks) and $E$ represents the edges (that correspond to the flow of data among the tasks). $G$ corresponds to the execution plan of a data flow, since it defines the exact execution order of the tasks.
\item $T=\{t_1, ..., t_n\}$ is a set of tasks\footnote{In the remainder of the paper, we will use the terms tasks, services and activities interchangeably.} of size $n$. Each flow task is responsible for one or both of the following: (i) reading or retrieving or storing data, and (ii) manipulating data.
\item Let $E=\{edge_1, ..., edge_{m}\}$ be a set of edges of size $m$. Each edge $edge_i, 1\leq i \leq m$ equals to an ordered pair $(t_j,t_k)$ denoting that task $t_j$ sends data to task $t_k$. $m \leq \frac{n(n-1)}{2}$; otherwise $G$ cannot be acyclic.

\item Let $PC=(T',D)$ be another directed acyclic graph, where $T' \subseteq T$. $D$ defines the precedence constraints (dependencies) that might exist between pairs of tasks in $T'$.
More formally, $D=\{d_1, ..., d_{l}\}$ is a set of $l$ ordered pairs: $d_i = (t_j,t_k), 1\leq i \leq l, ~ 1\leq j < k \leq n$, where each such pair denotes that $t_j$ must precede $t_k$ in any valid $G$. In other words, $G$ should contain a path from $t_j$ to $t_k$. This implies that if $D$ contains $(t_a,t_b)$ and $(t_b,t_c)$, it must also contain $(t_a,t_c)$. The $PC$ graph corresponds to a higher-level, non-executable view of a data flow, where the exact ordering of tasks is not defined; only a partial ordering is defined instead.

\item Two execution plans $G_1$ and $G_2$ that respect all the precedence constraints in $PC$ are termed as \emph{logically equivalent} flows.
\end{itemize}

In this work we initially focus on \emph{single-input single-output (SISO)} flows. A SISO data flow is defined as a flow $G$ that contains only one task with no incoming edges from another task and only one task with no outgoing edges.
The task with no incoming edges is termed as the \emph{source} task and the task with no outgoing edges is termed as the \emph{sink} task. In a \emph{SISO} flow, there is a dependency edge $d$ from the source task to any other non-sink task, and from all non-source tasks to the sink task. 

\begin{figure}[tb!]
\centering
\vspace{-10pt}
\includegraphics[width=0.8\textwidth]{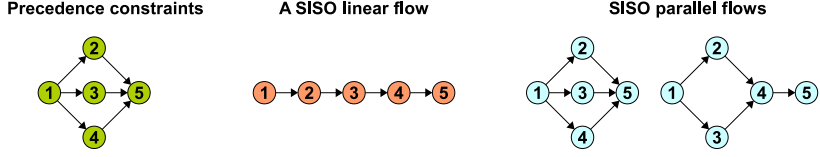}
\caption{Examples of two logically equivalent parallel execution plans of a \emph{SISO} linear conceptual data flow.}
\label{fig:linExampl}
\end{figure}

Examples of SISO flows are given in Figure \ref{fig:linExampl}. In the figure, we can see that a SISO flow can be executed both as a linear flow and as a parallel flow. In linear physical flows, $G$ has the form of a chain, and each non-source and non-sink task has exactly one incoming and one outgoing edge. In parallel physical flows, the output of a single task can be fed to multiple tasks in parallel. The linear flow and the parallel flows in the figure are logically equivalent flows. Because each SISO flow is logically equivalent to at least one linear $G$, we call SISO flows as \emph{logically (or conceptually) linear} flows.

Each task is further described as a triple $t_i = <c_i,sel_i,inp_i>$. In a dataflow, we assume that each task receives some data items as an input and outputs some other data items as a result. Following the database terminology, each data item is referred to as a {\it tuple}. The task elements are:

\begin{itemize}
\item {\it Cost} ($c_i$): we use $c_i$ = 1$/$$r_i, ~1\leq i \leq n$ as a metric of the time cost of each task, where $r_i$ is the maximum rate at which results of invocations can be obtained from the $i$-th task.

\item {\it Selectivity} ($sel_i$): it denotes the average number of returned data items per source tuple for the $i$-th service. For filtering operators, $sel_i < 1$, for data sources and operators that just manipulate the input $sel=1$, whereas, for operators that may produce more output records for each input record, $sel_i > 1$.
\item {\it Input} ($inp_i$): it denotes the size of the input of the $i$-th task $t_i$ in number of tuples per input data tuple. It depends on the product of the selectivities of the preceding tasks in the execution plan $G$.\footnote{Here, there is an implicit assumption that the selectivities are independent; if this is not the case, the product will be an arbitrarily erroneous approximation of the actual selectivity of the subplan before each task.} More formally, if $T_i^{prec}$ is the set of all preceding tasks of $t_i$ in $G$, $inp_i=\prod_{j=1}^{|T_i^{prec}|} sel_j$.
\item {\it Output} ($out_i$): The size of the output of the $i$-th task per source tuple can be easily derived from the above quantities, as it is equal to $inp_i sel_i$.

\end{itemize}

From the above quantities, and assuming that selectivities are independent, we can infer that $inp_i$ is the only task characteristic that depends on the position of $t_i$ in $G$; the cost and the selectivity of each task is independent of the exact $G$ that may include $t_i$.

{\bf Problem Statement:} Given a set of tasks $T$ with known cost and selectivity values, and a corresponding precedence constraint graph $PC$, we aim to find a valid $G$ that minimizes the \emph{sum cost metric (SCM)} per source tuple, defined as follows:
{\it SCM(G)}= $inp_i c_i + inp_2 c_2 + ... + inp_n c_n$.  The optimal plan is denoted as $P$.


Note that the input set of tuples are processed by all the tasks of the data flow, but typically, some of the input tuple attributes may not be required by every flow activity.
According to \cite{ZBML09}, the unnecessary tuple attributes just run through the flow, resembling an assembly-line model. The execution of a flow activity is not affected by the unnecessary attributes. This implies that the tasks of a flow have the ability to be reordered as long as the precedence constraints between the tasks are preserved.

\subsection{Problem Complexity}
In \cite{BMS05} it is proved that finding the optimal ordering of tasks is an $NP$-hard problem when (i) each flow task is characterized by its cost per input record and selectivity; (ii) the cost of each task is a linear function of the number of records processed and that number of records depends on the product of the selectivities of all preceding tasks (assuming independence of selectivities for simplicity); and (iii) the optimization criterion is the minimization of the sum of the costs of all tasks. All the above conditions hold for our case, so our problem is intractable. Moreover, in \cite{BMS05} it is discussed that \emph{``it is unlikely that any polynomial time algorithm can approximate the optimal plan to within a factor of $O(n^\theta)$"}, where $\theta$ is some positive constant. Note that if we modify the optimization criterion, e.g., to optimize the bottleneck cost metric or the critical path renders the problem tractable \cite{Sriv06,ABDR12}.
\section{Motivational Case Study in Kettle}
\label{pentaho}

\begin{figure}[b!]
\centering
\vspace{-10pt}
\includegraphics[width=1.0\textwidth]{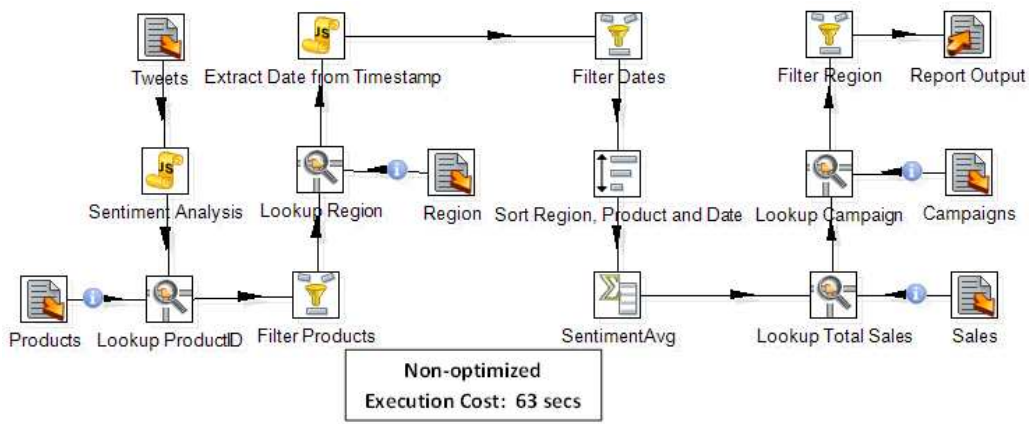}
\caption{A real-world analytic flow.}
\label{fig:nonOptimWFilter}
\end{figure}

In this section, we present the application of data flows in a real-world business tool, named as Pentaho Data Integration (PDI) (Kettle) \cite{Pentaho}, in order to highlight the impact of optimization proposals in the performance of a flow execution. We introduce a data flow (Figure \ref{fig:nonOptimWFilter}) that analyzes tags referring to products, which are retrieved from tweets in \emph{Twitter}, in order to compose a dynamic report that associates sales with marketing campaigns. In the following, we analyze the tasks of this data flow and of a flavour of it combined by details of the data set that the data flow process for the case study purposes.

As we observe this data flow has a single streaming source that outputs tweets on products and the flow accesses four other static sources through lookup operations. The initial streaming source task, called as \emph{Tweets}, of the flow consists of 1,000,000 records of tweets with attributes, such as product references, coordinates, timestamps etc. More specifically, the data flow is described as follows. When a tweet arrives as a timestamped string attribute ({\it tag}), the first task is to compute a single sentiment value in the range [-5 +5] for the product mentioned in the tweet (\emph{Sentiment Analysis}). Then, a lookup operation which maps product references in the tweet is performed (\emph{Lookup ProductID}) and after this a filter is applied in order to choose products with a specified range of product id values (\emph{Filter products}). The next task is also a lookup task which maps geographic information (latitude and longitude) in the tweet to a geographical region (\emph{Lookup Region}). In
the following, the task \emph{Extract date from timestamp} converts the tweet timestamp to a date and then, another filter is applied for choosing dates for a specific period of time (\emph{Filter Dates}). In order to implement the task \emph{SentimentAvg}, where the sentiment values are averaged over each region, product, and date, we first have to sort the values of region, product, and date by applying the task \emph{Sort Region, Product and Date}. The flow continues with other two lookup operations: the former maps the total sales of a product by the region, product and date (\emph{Lookup Total Sales}) and the latter maps campaigns of interest according to the results of total sales taken from the previous task (\emph{Lookup Campaign}). Finally, the user has the option to narrow down the report in order to focus on a specific region with the filtering task \emph{Filter Region}.

Additionally, there are four intermediate static sources, used as inputs in lookup operations, whose cost is embedded in the cost of the task where the static records are taken as inputs of the lookup task executions. The source task \emph{Products} has 100 records of product names and ids, while that \emph{Region} source task has 100 records of set of coordinates corresponding to a region name. Another source static task named \emph{Sales} consists of 4,000 sale details, such as the sold product name, the price, the quantity, the region where the product was sold etc., and the last one static source task, named \emph{Campaings}, has 500 campaign ids combined with the day that these campaigns begin, the region that will take place, but also the product ids that each campaign concern.

\begin{table}[tb!]
\centering \setlength\tabcolsep{4pt}
\scriptsize
\begin{tabular}{|c|c|c|c|}
\hline
\emph{ID} & \emph{\textbf{Flow Task}} & \emph{\textbf{Cost(secs)}} & \emph{\textbf{Selectivity}}\\ \hline\hline
1 & Tweets (data source) & 1.7 & 1 \\ \hline
2 & Sentiment Analysis & 4.5 & 1 \\ \hline
3 & Lookup ProductID & 5 & 1 \\ \hline
4 & Filter Products & 1.9 & 0.9 \\ \hline
5 & Lookup Region & 6.5 & 1 \\ \hline
6 & Extract Date from Timestamp & 19.4 & 1 \\ \hline
7 & Filter Dates & 2 & 0.2 \\ \hline
8 & Sort Region, Product and Date & 173 & 1 \\ \hline
9 & SentimentAvg & 10.3 & 0.1 \\ \hline
10 & Lookup Total Sales & 10.8 & 1 \\ \hline
11 & Lookup Campaign & 11.6 & 1 \\ \hline
12 & Filter Region & 2 & 0.22 \\ \hline
13 & Report Output & 1 & 1 \\ \hline
\end{tabular}
\caption{The cost and selectivities values.}
\label{tab:pentahoMetadata}
\vspace{-0.4cm}
\end{table}

Table \ref{tab:pentahoMetadata} shows the selectivity and cost values computed for a specific dataset of 1M records using a machine with an Intel Pentium G860 CPU and 4 GB of RAM. We can observe that the most expensive tasks are the grouping and lookup tasks, the cost of which is up to two orders of magnitude compared to the less expensive ones. Also, there are three filtering tasks, while the rest of the tasks do not modify the number of records (note that in general, selectivities may be higher than 1). In this data flow scenario, the selectivity values of the lookup and transformation tasks is 1, while the selectivity values corresponding to filtering and grouping tasks varies.

In Table \ref{tab:preCons} the precedence constraints that tasks have between them are presented, having in the left part of the arrows the tasks that must precede the tasks that are defined in the right part of the table. This data flow has 38\% precedence constraints, as they described in Table\ref{tab:preCons}, where a fully constrained flow with $n$ tasks and 100\% PCs  has $\frac{n(n-1)}{2}$ constraints and no equivalent ordering alternatives. In real data flow scenarios the preserving precedence constraints are approximately 30\% or even more, as the flows presented in \cite{RHHLN13}.

\begin{table}[tb!]
\centering \setlength\tabcolsep{1pt}
\caption{The precedence constraints of the data flow in Figure.}
\label{tab:tableRealStruct1}
\scriptsize
\begin{tabular}{|c c c|}
\hline
\multicolumn{3}{|c|}{\textbf{Precedence Constraints}}\tabularnewline\hline
$Sentiment Analysis$ & $\rightarrow$ & $SentimentAvg$ \\ \hline
$Lookup ProductID$ & $\rightarrow$ & $Filter products$ \\ \hline
$Lookup ProductID$ & $\rightarrow$ & $Sort Region, Product and Date$ \\ \hline
$Lookup ProductID$ & $\rightarrow$ & $Lookup Total Sales$ \\ \hline
$Lookup ProductID$ & $\rightarrow$ & $Lookup Campaign$ \\ \hline
$Lookup Region$ & $\rightarrow$ & $Sort Region, Product and Date$ \\ \hline
$Lookup Region$ & $\rightarrow$ & $Lookup Total Sales$ \\ \hline
$Lookup Region$ & $\rightarrow$ & $Lookup Campaign$ \\ \hline
$Lookup Region$ & $\rightarrow$ & $Filter Region$ \\ \hline
$Extract date from timestamp$ & $\rightarrow$ & $Filter Dates$ \\ \hline
$Extract date from timestamp$ & $\rightarrow$ & $Sort Region, Product and Date$ \\ \hline
$Extract date from timestamp$ & $\rightarrow$ & $Lookup Total Sales$ \\ \hline
$Extract date from timestamp$ & $\rightarrow$ & $Lookup Campaign$ \\ \hline
$Sort Region, Product and Date$ & $\rightarrow$ & $SentimentAvg$ \\ \hline
\end{tabular}
\label{tab:preCons}
\end{table}

\begin{figure}[b!]
\centering
\vspace{-10pt}
\includegraphics[width=1.0\textwidth]{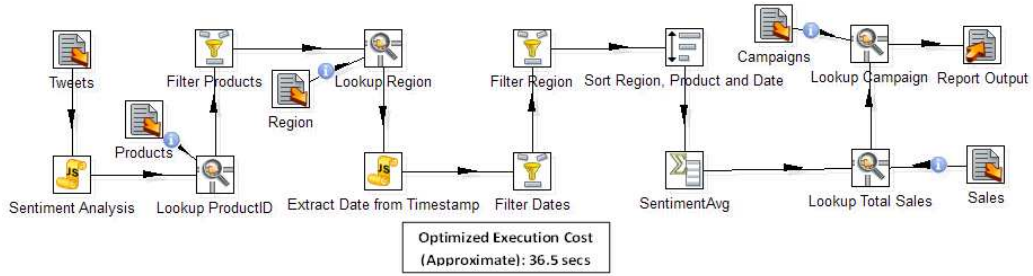}
\caption{The optimized plan by a heuristic algorithm of the flow in Figure \ref{fig:nonOptimWFilter}.}
\label{fig:optimSwapWFilter}
\end{figure}

\begin{figure}[b!]
\centering
\vspace{-10pt}
\includegraphics[width=1.0\textwidth]{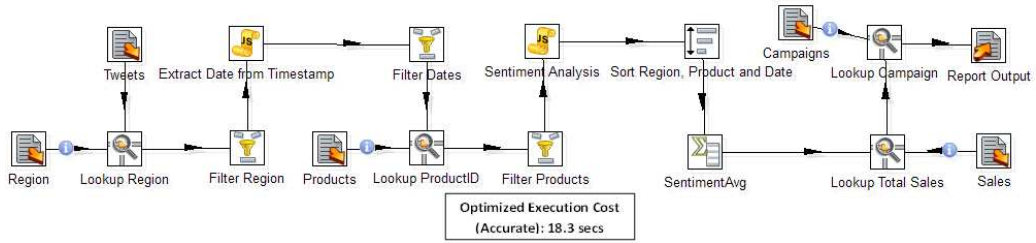}
\caption{The optimized plan by an exhaustive algorithm of Figure of the flow in \ref{fig:nonOptimWFilter}.}
\label{fig:optimTsWFilter}
\end{figure}

A straight-forward implementation is shown in Figure \ref{fig:nonOptimWFilter}. Then, we applied best-performing approximate heuristic to date, which is proposed in \cite{671SVS05}. The optimized plan is illustrated in Figure \ref{fig:optimSwapWFilter}. In that case, the performance improvement from the initial non-optimized flow is 42\% from 63 to 36.5 seconds.

Similar to the previous optimization, we applied our exhaustive solution to the flow of Figure \ref{fig:nonOptimWFilter} in order to find the optimal flow execution cost. In Figure \ref{fig:optimTsWFilter} the optimal plan of the initial data flow is depicted. In this case, the exhaustive optimization methodology transposes the filtering task \emph{Filter Region}, which at the initial design has been placed at the end as a final optional step,  at the very beginning for this specific flow due to the metadata in Table \ref{tab:pentahoMetadata}. A less obvious optimization is to move the pair of date extraction and filtering tasks upstream although the former is expensive and not filtering. The execution cost of this optimized plan is 18.3 and results to a plan that is 3 times better than initial non-optimized. Both of the mentioned optimization methodologies are analyzed in the following sections.

This is a representative example of a real manually designed data flow that exhibits significantly suboptimal behavior. In general, we can draw two observations. Firstly, optimal solutions may yield lower execution costs by several factors. A second equally important observation is that even in simple cases like the one examined here, existing heuristics may fail to closely approximate the optimal solution and generate the plan in Figure \ref{fig:optimTsWFilter}. The main reason in this example is that the approximate solution performs greedy swaps of adjacent activities; however the region filter cannot move earlier unless the campaign lookup task is moved earlier as well, an action that a greedy algorithm cannot cover.

\section{Accurate Algorithms for Linear Execution Plans}
\label{accurate}
In this section, we present three accurate algorithms for reordering \emph{SISO} data flows in order to generate an optimal execution plan. The algorithms are based on  backtracking, dynamic programming and generation of all topological sortings, respectively. Our main novelty here is that we examine a topological sorting-based algorithm, despite its worst-case complexity. Counter-intuitively, as we show in the evaluation, the algorithm is practical even for large $n$, when there are many precedence constraints and, in general, can scale better than the two other options. However, still, it cannot be applied to arbitrary flows of medium or large size.

\subsection{Backtracking}
\label{subsec:backtrack}

The \emph{Backtracking} algorithm  finds all the possible execution plans generated after reordering the tasks of a given data flow preserving the precedence constraints. The algorithm enumerates all the valid sub-flow plans after applying a set of recursive calls on these sub-flows until generating all the possible data flow plans. It backtracks when a placement of a task in a specific position violates the precedence constraints. The algorithm is proposed for flow optimization in \cite{HPS+12}.

{\it Complexity:} The worst case time complexity of \emph{Backtracking} is factorial (i.e., $O(n!)$), since, if there are no dependencies, all orderings will be examined in a brute force manner.

\subsection{Dynamic programming}
\label{DP}
This algorithm is extensively used as part of the System R-type of query optimization to produce (linear) join orderings \cite{SAC+79}.
The rationale of the dynamic programming algorithm (termed as \emph{DP} henceforth) for data flows remains the same, that is to calculate the cost of task subsets of size $n$ based on subsets of size $n-1$. For each of these subsets, we keep only the optimal solutions, which are valid with regards to the precedence constraints. Specifically, the \emph{DP} algorithm considers each flow of size $n$ as a flow of $(n-1)$ tasks followed by the $n$th task; the key point is that the former part is the optimal subset of size $n-1$, which has been found from previous step; then the  algorithm exhaustively examines which of the $n$ flow tasks is the one that, when added at the end, yields an optimal subplan of size $n$.
For example, the algorithm starts by calculating subsets that consist of only one task $\{t_1\}$, then $\{t_2\}$, $\{t_3\}$ and so on. In a similar way, in the second step, it examines subsets containing two tasks, i.e., $\{t_1,t_2\}$, $\{t_1,t_3\}$ and so on, until it examines the complete flow $\{t_1,t_2, ...,t_n\}$. The number of the optimal (non-empty) subsets of a flow is equal to $2^{n}-1$. More details, along with pseudocode and an example are provided in \ref{sec:app-dp}.

{\it Complexity:} The time complexity is $O(n^2 2^n)$. This is because we examine all subsets of $n$ tasks, which are $O(2^n)$. For each subset, which is up to size $O(n)$, we examine whether each element can be placed at the end of the subplan. Each such check involves testing whether any of the rest $n-1$ tasks violate a precedence constraint, when placed before the $n$-th task. Overall, for each element, we make $O(n)$ comparisons. So, the overall time complexity is $O(2^n)O(n)O(n) = O(n^2 2^n) $. The space complexity is derived by the size of the auxiliary data structures employed. We use three vectors of size $2^{n}-1$ as explained in \ref{sec:app-dp}, the one of which stores elements of size $O(n)$. So the space complexity is $O(n2^n)$.


\subsection{Topological sorting}
The \emph{TopSort} algorithm is a topological sorting algorithm based on \cite{VR81}, which finds all the possible topological sortings given a partial ordering of a finite set; in our case the partial ordering is due to the precedence constraints. The reason behind using this algorithm is that it (implicitly) prunes invalid plans very efficiently and it generates a new plan based on a previous plan after performing a minimal change.
For the purposes of this work, we adapted the topological sorting algorithm  in order to generate all the possible execution plans of a data flow and detect the execution plan with the minimum cost. The algorithm assumes that it can receive as input a valid task  permutation $t_1 \rightarrow t_2 \rightarrow t_3 \rightarrow ... \rightarrow t_n$, which is trivial since it can be done in linear time. We generate all other valid execution plans by applying cyclic rotations and swapping adjacent tasks.

Firstly, the process of generating all the valid flow execution plans begins with the topological sorting of the $n-1$ tasks $t_2 \rightarrow t_3 \rightarrow ... \rightarrow t_n$ of the flow. Based on this partial sorting, we generate all the valid orderings of the $t_1 \rightarrow t_2 \rightarrow t_3 \rightarrow ... \rightarrow t_n$ plan. Specifically, in the first stage of the algorithm the task $t_1$ is placed on the left part of the partial plan $t_2 \rightarrow t_3 \rightarrow ... \rightarrow t_n$ and in the next steps of this stage, we swap it with the tasks on its right, while the tasks of the partial plan maintain their relative position. The $t_1$ stops moving when such a swap violates a precedence constraint. Then, as the task $t_1$ cannot be further transposed, the second stage of algorithm begins with a right-cyclic rotation of another partial plan consisted of $t_1$ and all the tasks that precede it, which means all the tasks which are positioned to its left. In this way, $t_1$ is placed to its
initial position. Similarly, we generate all the topological sortings of $t_2 \rightarrow t_3 \rightarrow ... \rightarrow t_n$, $t_3 \rightarrow t_4 \rightarrow ... \rightarrow t_n$ and so on. For example, the topological sorting of $t_4 \rightarrow t_5 \rightarrow ... \rightarrow t_n$ partial plan will be generated with the transpositions of task $t_4$. For each generated plan, we estimate the total execution cost and finally, we choose the flow execution plan with the best performance. A pseudocode, a brief example and further details are in \ref{sec:app-ts}.

{\it Complexity}: Since the algorithm checks all the permutations the time complexity is $O(n!)$ in the worst case. However, compared to other algorithms that produce all topological sortings, it is more efficient \cite{VR81}.
The space complexity is $O(n)$ because only one plan is stored in main memory at any point of execution.

\section{Approximate Algorithms for Linear Execution Plans}
\label{approximate}
Due to the high complexity of the problem in hand, we need to develop approximate solutions for the generic case. This section consists of two parts: we first present existing solutions including straightforward extensions of existing proposals that are applicable to our problem, and then we present our main novelty with regards to approximate optimization of linear data flows.
As will be shown in the evaluation, there is a significant gap in the performance between optimal solutions and existing approximate algorithms, and our proposal fills that gap.\footnote{An initial introduction of the existing algorithms has appeared in \cite{KG13}.}

\subsection{Existing Solutions}
Here we present four algorithms, which reflect the current state-of-the-art in task re-ordering in linear flows. Implementation details and examples are provided in \ref{sec:app-appr}.

\subsubsection{Swap}

The $Swap$ algorithm starts with a random valid execution plan. Such a plan is trivial to be computed in linear time through a single topological ordering of $PC$. The algorithm then
compares the cost of the existing execution plan against the cost of the transformed plan, if we swap two adjacent tasks provided that the constraints are always satisfied. We perform this check for every pair of adjacent tasks and we repeat until no changes occur. {\it Swap} is equivalent to the proposal in \cite{671SVS05} when only task re-ordering is allowed.  The complexity of the \emph{Swap} algorithm is $O(n^2)$ because we can repeat at most $n$ times, and each iteration has $O(n)$ complexity. The space complexity is linear ($O(n)$), equal to the complexity needed to store a single plan.

In order to prove that {\it Swap} is approximate, it is adequate to provide at least one example that the algorithm fails to yield the optimal solution. Assume a flow, which has three inner tasks (i.e., tasks other than the source and sink ones), each with cost equal to 1 and selectivities 1, 1.1, and 0.5 respectively. There is also a precedence constraint between tasks 2 and 3. If the initial plan is $t_1 \rightarrow t_2 \rightarrow t_3$, then its $SCM=1+1 + 1.1 = 3.1$. However, the optimal plan is $t_2 \rightarrow t_3 \rightarrow t_1$ with $SCM=1+1.1 + 0.55 = 2.65$. \emph{Swap} cannot produce that plan because it cannot perform transpositions that initially produce worse plans, but eventually lead to better solutions, such as the swap of tasks $t_1$ and $t_2$.

\subsubsection{GreedyI and GreedyII}

\emph{GreedyI} starts with an empty plan and in each step, it adds the activity with the maximum value of $(1-sel_{i})/(c_i)$, provided that it meets the precedence constraints. In the first step, the source task is chosen as the only eligible one. It bears similarities with the {\it Chain} algorithm in \cite{YLUG99}, although the latter algorithm was proposed for a different problem and appends the activity that minimizes $c_i$.
The time complexity of \emph{Greedy} algorithm is $O(n^2)$ because it consists of $n$ steps, where in each step $O(n)$ checks are performed to find the most efficient and valid task to append. With the help of appropriate data structures, the complexity can drop to $O(nlogn)$.

Similarly to \emph{Swap}, it may miss the optimal solution. For example, in the example with the three tasks of cost 1 and selectivities 1, 1.1 and 0.5,  and a precedence constraint between $t_2$ and $t_3$, \emph{GreedyI} will first append $t_1$, then $t_2$ and last $t_3$, which is not the best possible plan as explained earlier.

Another greedy algorithm is \emph{GreedyII} \cite{KK10}. The rationale of \emph{GreedyII} is similar to \emph{GreedyI} apart from the fact that the construction of the optimized execution plan is right-to-left (i.e., from the sink to the source).

\subsubsection{Partition}
$Partition$ forms clusters with activities by taking into consideration their eligibility. Specifically, each cluster consists of activities that their prerequisites have been considered in previous clusters. After building the clusters, each cluster is optimized separately by checking each permutation of cluster tasks. Similar to \emph{GreedyI}, it was first proposed for data integration systems, and the details are given in \cite{YLUG99}.
$Partition$ runs in $O(n!)$ time in the worst case because, if there are no precedence constraints, it checks all permutations of a partition of size $n$. In general, its complexity is $O(k!)$, where $k$ is the size of the largest cluster, and thus is inapplicable if a cluster contains more than a dozen of tasks. As in the previous optimality examples with the three tasks, it is easy to verify that it cannot find the optimal plan. In the first step, it forms a cluster with tasks $t_1$ and $t_2$ and decides to place $t_1$ before $t_2$ because is yields a better subplan.

\subsection{Algorithms based on rank ordering}
\label{sec:ro}


\begin{figure}[tb!]
\centering
\begin{minipage}{0.45\textwidth}
\includegraphics[width=1.1\textwidth]{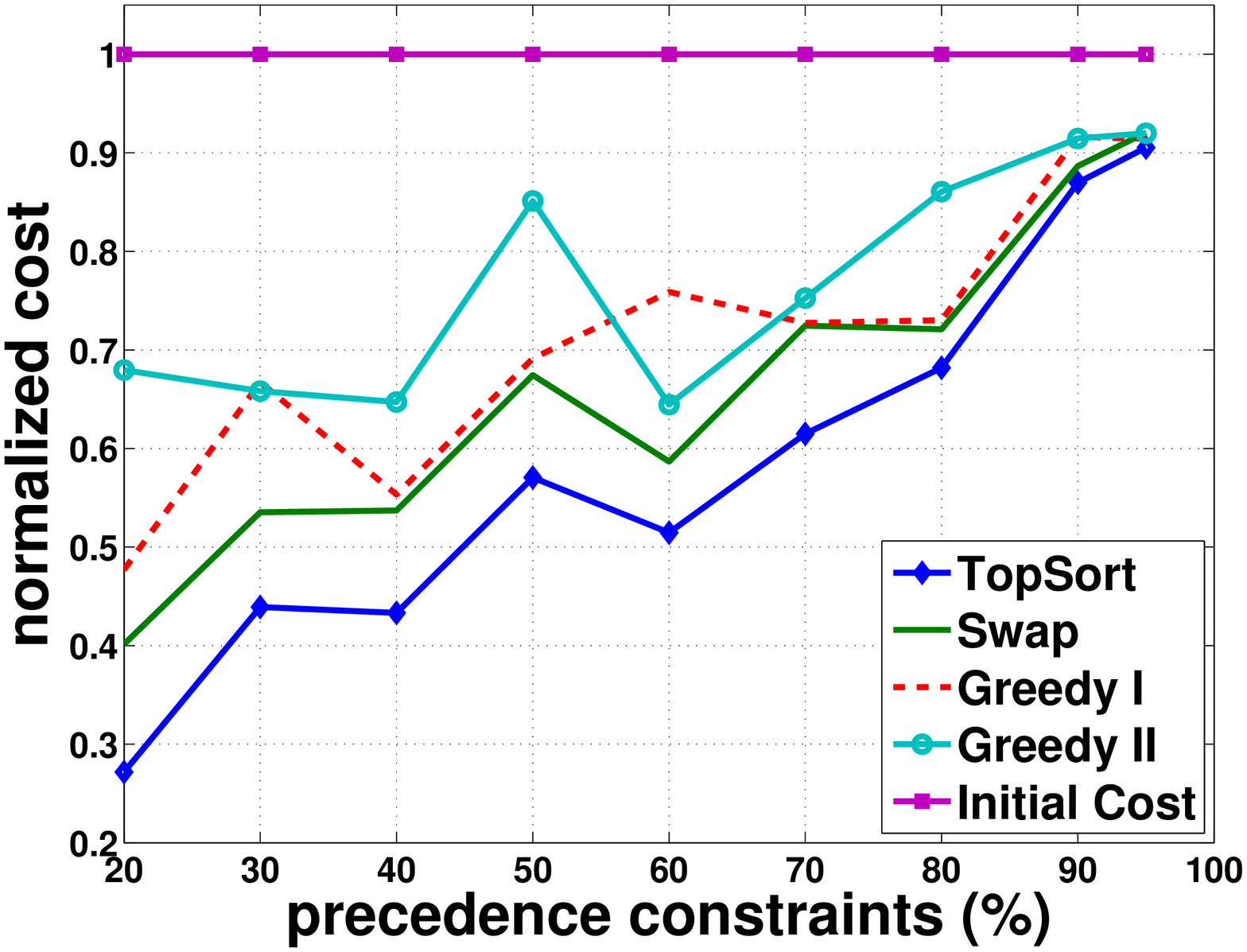}
\end{minipage}
\begin{minipage}{0.45\textwidth}
\includegraphics[width=1.1\textwidth]{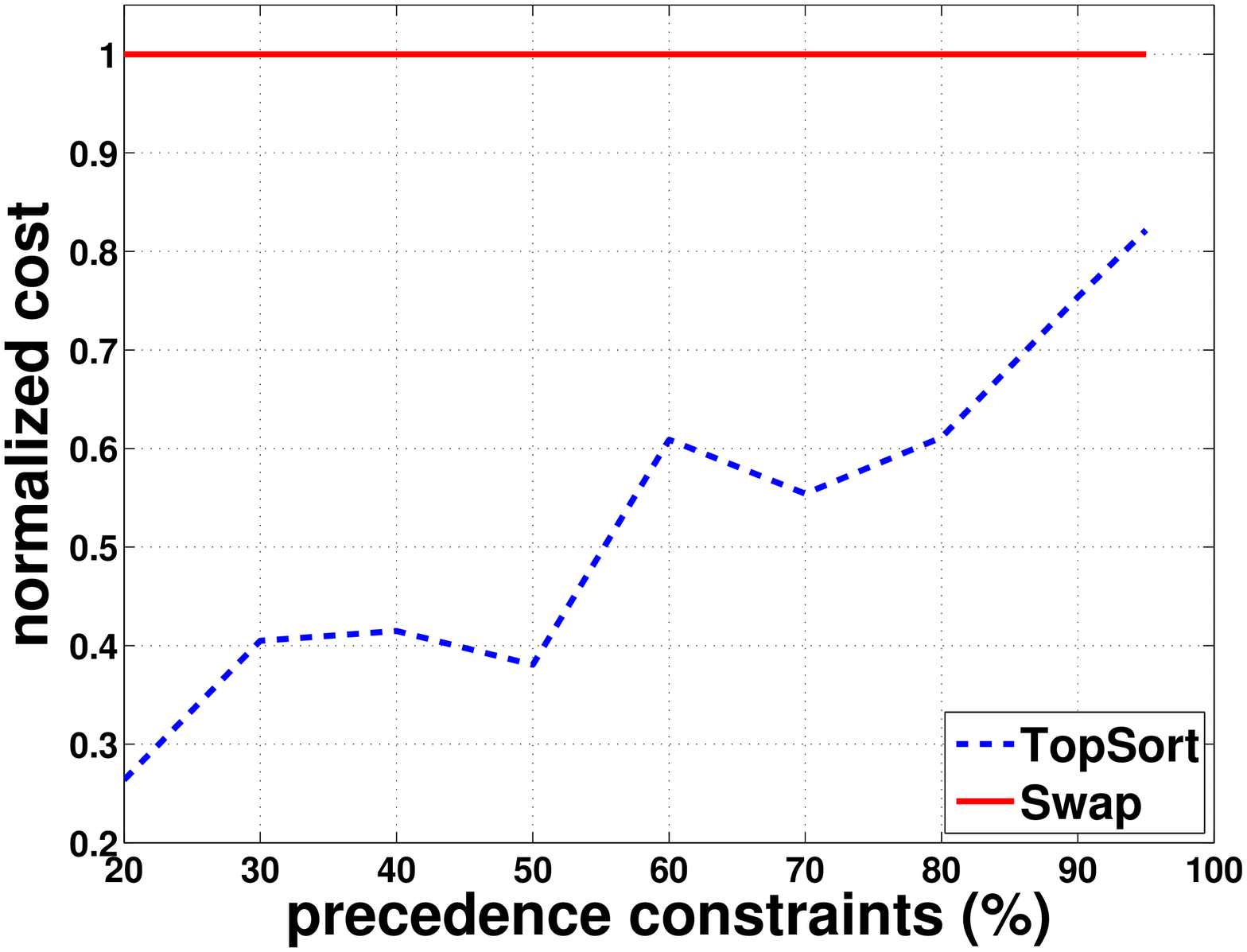}
\end{minipage}
\caption{Average (left) and maximum (right) improvements of exhaustive solutions}
\label{fig:heuristic}
\vspace{-0.3cm}
\end{figure}

The motivation behind our proposal is that the approximate solutions discussed previously deviate significantly from the optimal orderings. To prove this, we conduct experiments with small flows, where applying an exhaustive technique to obtain the optimal plan is feasible. More specifically,
in Figure \ref{fig:heuristic} (left), we examine 100 randomly generated data flows consisting of 15 tasks with $cost \in [1,100]$, $sel \in (0,2]$ and 20\%-95\% precedence constraints. The results show that the performance improvement derived by the application of an accurate algorithm is high; see that \emph{TopSort} algorithm can have up to 57\% better performance improvement compared to a random initial flow that just respects the precedence constraints. In general, \emph{Swap} seems to be the heuristic algorithm with the best performance improvement on average. In Figure \ref{fig:heuristic}(right), the maximum normalized difference between \emph{Swap} and \emph{TopSort} algorithms is presented. As we can observe, there are cases where the \emph{TopSort} algorithm has 74\% better performance improvement than the best heuristic. These findings highlight the need for proposing new approximate optimization methodologies, in order to provide more near-optimal flow execution plans.

\begin{algorithm}[tb!]
\caption{Rank ordering based high-level algorithm}
\begin{algorithmic}[1]
\REQUIRE A set of n tasks, T=\{$t_1$, ..., $t_n$\}  and the PC graph
\ENSURE A directed acyclic graph P representing the optimal plan
\STATE Pre-processing phase
\STATE Apply KBZ algorithm
\STATE Post-processing phase
\end{algorithmic}
\label{alg:ROgeneric}
\end{algorithm}

To fulfill this need, we propose a set of rank ordering-based approximate algorithms and we analyze them in this section. We build upon the join ordering algorithms proposed for query optimization in \cite{Ibaraki84,KBZ86}, which will be referred to as \emph{KBZ}. This algorithm leverages the rank value of each task defined as  $\frac {1-sel_i}{c_i}$ and the dependencies among tasks. Our solutions can be described at a high-level as shown in Algorithm \ref{alg:ROgeneric}. The main novelty is how to preprocess the flow, so that \emph{KBZ} becomes applicable. Also, we post-process the result of the \emph{KBZ} algorithm in order either to guarantee validity or to further improve the intermediate results. There are many options regarding how these two phases can be performed and here we present three concrete suggestions, which constitute the novelty of this section (examples are shown in \ref{sec:app-ros}).

\subsubsection{KBZ}

The \emph{KBZ} algorithm, which was proposed in \cite{KBZ86}, is a seminal query optimization algorithm for join ordering. This algorithm considers only a specific form of precedence constraints, namely those representable as a rooted tree. The rationale of this algorithm is to order tasks according to their rank value. In the case that this is not possible due to the defined precedence constraints, the tasks are merged and the rank values are updated accordingly. The fact that \emph{KBZ} algorithm allows only tree-shaped precedence constraint graphs implies that there should be no task with more than one independent prerequisite activity, and in such data flow scenarios, the percentage of precedence constraints is very low and decreases more with the number of tasks (e.g., less than 10\% for a 100-node flow). Both of these cases do not occur frequently in practice. The time complexity of \emph{KBZ} algorithm is $O(n^2)$.

\subsubsection{RO-I}
In our first proposal, called \emph{RO-I},  the pre-processing phase ensures the transformation  of the PC graph into a tree-shaped one. This is done by removing incoming edges with no maximum rank, if a task has more than one incoming edge. This allows KBZ to run but may produce invalid flow orderings. To fix that, we employ a post-processing phase where any resulting PC violations are resolved by moving tasks upstream if needed as prerequisites for other tasks placed earlier.

The worst case complexity of the pre-processing phase  is $O(n^2)$ because we remove up to $n-1$ incoming edges ($O(n)$ complexity) from each task and we repeat this for $n-1$ tasks of the flow. Additionally, in the  post-processing step, we check, for each of the $n$ tasks, if any of the preceding tasks violates the precedence constraints. There can be up to $n-1$ preceding tasks in a flow ordering.
So, in the worst case, the complexity is $O(n^2)$. However, in practice the average time complexity is much lower for both phases.

%

\subsubsection{RO-II}
\begin{figure}[tb!]
\centering
\vspace{-10pt}
\includegraphics[width=0.3\textwidth]{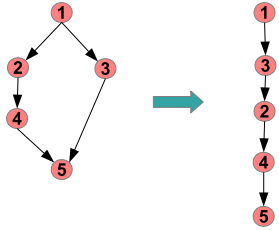}
\caption{A merging example.}
\label{fig:ROIImerge}
\end{figure}

The \emph{RO-II} algorithm follows a different approach in order to render \emph{KBZ} applicable. In the pre-processing phase, this approximate algorithm first detects paths in the precedence constraint graph that share an intermediate source and sink. Then it merges them to a single path based on their rank values. When there are multiple such paths, we start merging from the most upstream ones and when there are nested paths, we start merging from the innermost ones. In that way, all precedence constraints are preserved at the expense of implicitly examining fewer re-orderings. An example is shown in Figure \ref{fig:ROIImerge}. In that example,  after the merging procedure we enforce more precedence constraints than the original ones, so that the task $t_3$ must precede not only task $t_5$ but also tasks $t_2$ and $t_4$. In other words, the merging process imposes more restrictions on the possible re-orderings. As such, these local optimizations may still deviate from a globally optimal solution
significantly in the average case. \emph{RO-II} does not require any post-processing because its result is always valid.

\emph{RO-II}, as will be shown in the evaluation section, in general behaves better than \emph{RO-I}. However, in some cases \emph{RO-II}'s performance is much worse and an example is in \ref{sec:app-ros}. Also, in the case of \emph{RO-II}, the time complexity remains $O(n^2)$ because for each merge process we consider at most $O(n)$ flow tasks and we repeat this for all the possible merge processes that can be up to $n$.

\begin{algorithm}[tbh!]
\caption{RO-III}
\begin{algorithmic}[1]
\REQUIRE A set of n tasks, T=\{$t_1$, ..., $t_n$\} \\
         A directed acyclic graph PC with precedence constraints\\
         Optimized plan P as a directed acyclic graph returned by RO-II
\ENSURE A directed acyclic graph P representing the optimal plan
\REPEAT
\STATE \COMMENT{$k$ is the maximum subplan size considered}
\FOR {i=1:k}
\FOR {s=1:n-i}
\FOR {t=s+i:n}
\STATE consider moving subplan of size $i$ starting from the $s^{th}$ task after the $t^{th}$ task
\ENDFOR
\ENDFOR
\ENDFOR
\UNTIL {no changes applied}
\end{algorithmic}
\label{alg:ROIII}
\end{algorithm}

\subsubsection{RO-III}
After the evaluation of the proposed \emph{RO-I} and \emph{RO-II} algorithms, we isolated data flow cases that were not near-optimal. For example, the \emph{RO-II} was not able to reorder a filtering task in an earlier stage of the flow, even when was not restricted by precedence constraints, in order to reduce the data that the flow will process. To fill this gap, we propose \emph{RO-III} to support the efficient optimization of such data flow cases.
The \emph{RO-III} algorithm, presented in Algorithm \ref{alg:ROIII}, tackles the limitations of \emph{RO-II} with the help of a post-processing phase that we introduce. Specifically, we apply the \emph{RO-II} algorithm in order to produce an intermediate execution plan, and then we examine several
transpositions. More specifically, we check all the possible transpositions of each sub-flow of size from 1 to $k$ tasks in the plan. The checks are applied from the left to the right. In this way, we address the problem of a task being ``trapped" in a suboptimal place upstream in the flow execution due to the additional implicit constraints introduced by \emph{RO-II} (see the transposition of $t_7$ in Figure \ref{fig:ROIIIexample1} in \ref{sec:app-ros}). This process is described by the 3 nested $for$ loops in Algorithm \ref{alg:ROIII} and is repeated until there are no changes in the flow plan. The reason we repeat it is because each applied transposition may enable further valid transpositions that were not initially possible.

The post-processing phase of the \emph{RO-III} algorithm has $O(kn^2)$ complexity, which is derived by the maximum number each of the three inner loops can execute. The \emph{repeat} process in theory can execute up to $n$ times, but in practice, even for large flows, there is no change after 3 times. In all experiments, we set $k$ to 5.

\section{Parallel Optimization Solutions}
\label{sec:parallel}

This section focuses on the advantages of parallel execution plans. As we have explained in Section \ref{sec:background}, in a parallel physical flow each single task can have multiple outgoing edges, which implies that the output of such a task is fed, as input, to multiple tasks. In the right part of Figure \ref{fig:linExampl}, we observe that a single task may have not only multiple outgoing edges, but also multiple ingoing edges. In this case, a single task receive as input data the output of multiple tasks. This is in line with the \emph{AND-Join} workflow pattern as presented in \cite{AHKB03}, where the outgoing edge of multiple tasks that are executed in parallel converge into a single task.

This case can be considered as a merge-split process, which in software tools such as PDI can be implemented by incorporating a \emph{merge join} process. As such, merging multiple input streams incurs an extra execution cost. To assess this cost, we evaluated parallel data flows that were executed with the PDI tool. The conclusion was that the merge task cost has a small effect on the total flow execution cost; in other words, the merge task is similar to an additional lightweight activity. Additionally, the size of the input ($inp_i$) of a task $t_i$, which receives more than one incoming edge is defined similarly to the tasks with only one incoming edge, i.e., by computing the  the product of the selectivity values of the preceding tasks as we have described in Section \ref{sec:background}.


\begin{figure}[tb!]
\centering
\vspace{-10pt}
\includegraphics[width=0.7\textwidth]{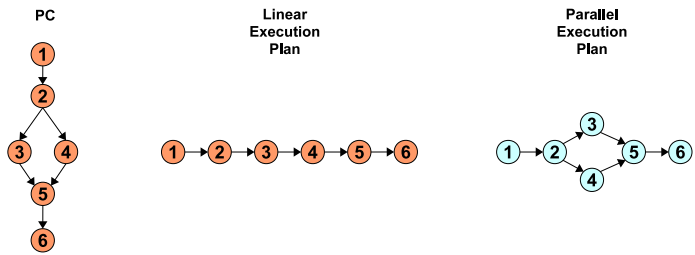}
\caption{An example of flow parallel execution.}
\label{fig:parallelTheory}
\end{figure}

Let us now analyze when the parallel flow execution may be beneficial through a theoretical example. Let us consider two subsequent tasks $t_3$ and $t_4$ illustrated in Figure \ref{fig:parallelTheory}, which do not have precedence constraints between them and an extra cost of the merge process that will be denoted as $mc$. In this figure, we show two alternative plans, a linear one (in the middle) and a parallel one (on the right). The \emph{SCM} values of the two alternatives vary only with respect to activities $t_4$ and $t_5$. We distinguish between the following four cases (using a superscript to differentiate the inputs in the two cases):

\begin{itemize}
\item Case I: $sel_3 \leq 1$ and $sel_{4} \leq 1$. The linear execution cost is lower than the parallel execution cost, because (i) $inp_4^{linear}  c_4  < inp_4^{parallel} c_4$ as $inp_4^{linear} = sel_3 inp_4^{parallel}$ and $sel_3 <1$, and (ii) $inp_5^{linear} c_5 < inp_5^{parallel} (c_5 + mc)$ due to the extra merge cost of the parallel version and given that $inp_5^{linear} = inp_5^{parallel}$. So, in that case, parallelism is not beneficial.
\item Case II: $sel_3 \leq 1$ and $sel_{4} > 1$. Similar with the Case I, the linear execution of the flow is more beneficial than the parallel; note that the selectivity value $sel_4$ does not affect the previous statements.
\item Case III: $sel_3 > 1$ and $sel_{4} > 1$. If $mc = 0$, the parallel execution results in better performance than the linear execution. In that case $inp_5^{linear}c_5 = inp_5^{parallel} (c_5 + mc)$. Because of the fact that $sel_3>1$, we deduce that $inp_4^{linear} c_4 > inp_4^{parallel} c_4$. In the generic case where $mc > 0$, we need to compute the estimated costs in order to verify which option is more beneficial, but we expect that, for small $mc$ values, the parallel execution to outperform.
\item Case IV: $sel_3 > 1$ and $sel_{4} \leq 1$. Following the rationale of the previous case, there is no clear winner between the two executions shown in Figure \ref{fig:parallelTheory}. However, an optimized linear plan will place $t_4$ before $t_3$ thus corresponding to Case I, where the (new) linear plan is better than the parallel one.
\end{itemize}

As in the previous section, we first describe a simple extension to an existing solution for ordering web services described in \cite{Sriv06}. Then we propose a novel post-processing step that applies to any of the solutions in the previous section in order to render their output plans into parallel ones. Our solution leverages and generalizes the analysis above, and based on the findings of Case III, it parallelizes tasks with selectivity higher than 1.

\subsection{PGreedyI and PGreedyII}

The \emph{PGreedyI} optimization algorithm has the distinctive feature of generating parallel flow execution plans. The rationale of the \emph{PGreedyI} is to order the flow tasks in such a way that the amount of data that is received by the tasks with selectivity value $>$ 1 is reduced by pushing the selective flow tasks (filtering tasks) in an earlier stage of the flow to prune the input dataset. Based on the selectivity values, the optimal execution plan may dispatch the output of a task to multiple other tasks in parallel, or place them in a sequence. Specifically, the flow tasks having selectivity value $>$ 1 are candidates for parallel execution in a flow. To this end, we employ the algorithm in \cite{Sriv06} for generating parallel flow execution plans. The detailed description of the \emph{PGreedyI} algorithm is presented in \ref{sec:app-pgreedy}.

A weak point of \emph{PGreedyI} is that, in each step, it tries to find the task that has the minimum cost without considering the implications for the next tasks (e.g., due to high selectivity).
The second flavour, \emph{PGreedyII}, chooses not the activity with the less cost but the activity with the highest rank value; in this way we penalize tasks that have low cost but high selectivity, which can yield lower \emph{SCM} values for the overall plan. Both algorithms have time compexity in $O(n^5)$ in the worst case, as explained in \cite{Sriv06}.

\subsection{Executing SISO flows in parallel}

\begin{algorithm}[tbh!]
\caption{Post-process step for parallel SISO flows}
\begin{algorithmic}[1]
\REQUIRE An optimized linear plan P=$\{t_1 \rightarrow ... \rightarrow t_n\}$ \\
         ~~~~~~~~~A directed acyclic graph PC with precedence constraints \\
\ENSURE A directed acyclic graph P representing the optimal parallel plan
\STATE i=1
\WHILE {$i<n$}
\STATE j=i+1
\WHILE {$sel_{P(j)} > 1$}
\STATE  Delete the edge between the tasks $t_{P(j-1)} \rightarrow t_{P(j)}$ from $P$
\IF {$t_{P(j)}$ is not predecessor in PC for no task in $t_{i+1} \dots t_{j-1}$}
\STATE Connect the edge between the tasks (i)~$t_{P(i)}$ and (ii) $t_{P(j)}$, i.e., create the edge $t_{P(i)}\rightarrow t_{P(j)}$ in $P$
\ELSE
\STATE Connect in $P$ the edge between (i) all the preceding tasks in PC with no outgoing edges in $P$ and (ii)~ $t_{P(j)}$
\ENDIF
\STATE $j=j+1$
\ENDWHILE
\STATE Connect in $P$ the edge between (i) all the tasks $t_{P(i+1)} \dots t_{P(j-1)}$ with no outgoing edges in $P$ and (ii)~ $t_{P(j)}$
\STATE i=j
\ENDWHILE
\end{algorithmic}
\label{alg:pSISO}
\end{algorithm}

\begin{figure}[tb!]
\centering
\vspace{-10pt}
\includegraphics[width=0.7\textwidth]{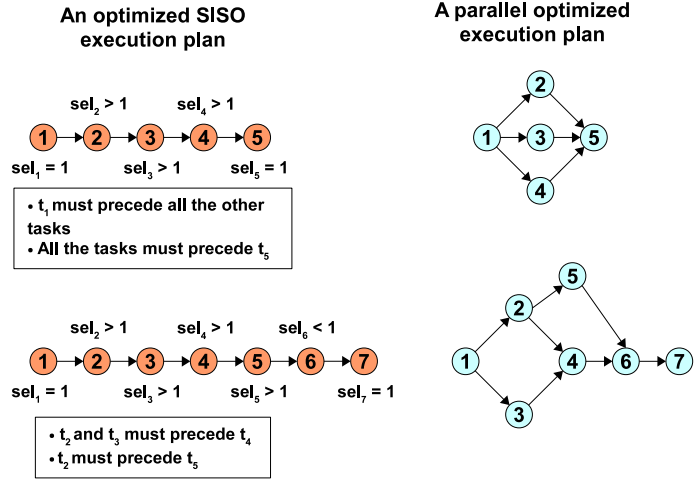}
\caption{Example of executing SISO flows in parallel.}
\label{fig:parallelExample}
\end{figure}

In order to exploit the advantages of the proposed optimization techniques, in Algorithm \ref{alg:pSISO}, we introduce a post-process phase for executing data flows in parallel. To this end, after the generation of an optimized linear execution plan, we apply a post-process step that restructures the flow in a way that subsequent tasks having selectivity greater than 1 to be executed in parallel if this does not incur violations of the precedence constraints. This post=process step can be applied to any optimization algorithm that produces a linear ordering.

An example is presented in Figure \ref{fig:parallelExample}, where in the upper flow scenario, we choose to parallelize the tasks $t_2$, $t_3$ and $t_4$, while in the flow case that is depicted in the bottom of the figure, we execute parallel only the tasks $t_2$ and $t_3$ and not $t_4$, because of the precedence constraints. Then, $t_5$ is appended after $t_2$ because of the constraints and is executed in parallel with $t_4$. As the task $t_6$ has selectivity value $<1$, it is not executed in parallel with any other task.

The complexity is $O(n^2)$. The parallelization of each task is examined at most once, and each such case the preceding tasks need to be checked, the number of which cannot exceed $n$.

\section{Extensions to MIMO flows}
\label{sec:mimo}

\begin{figure}[tb!]
\centering
\begin{minipage}{0.45\textwidth}
\includegraphics[width=1.1\textwidth]{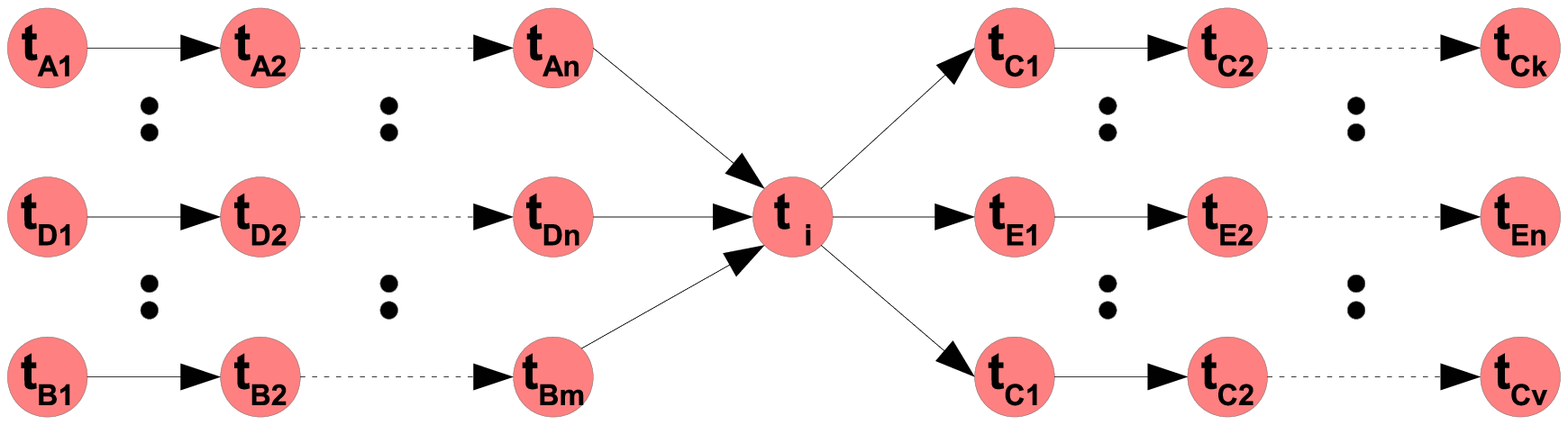}
\end{minipage}
\hfill
\begin{minipage}{0.45\textwidth}
\includegraphics[width=1.1\textwidth]{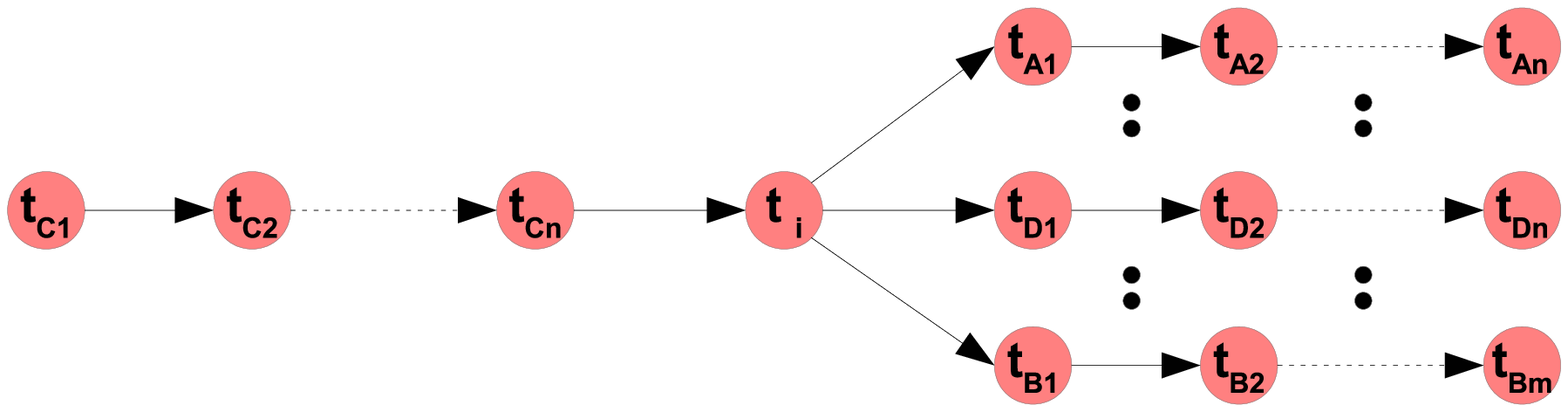}
\end{minipage}
\caption{Example MIMO data flows of type butterfly (left) and fork (right).}
\label{fig:MIMOexamples}
\vspace{-0.3cm}
\end{figure}

\begin{algorithm}[tbh!]
\caption{Optimization of MIMO flows}
\begin{algorithmic}[1]
\REPEAT
\STATE Extract SISO segments
\FOR {all SISO segments}
\STATE Optimize SISO segments
\ENDFOR
\STATE Apply factorize/distribute optimization thus modifying the SISO segments
\UNTIL{no changes}
\end{algorithmic}
\label{alg:MIMO}
\end{algorithm}

So far we have discussed the case with a single source and a single sink task, but arbitrary \emph{multiple-input multiple-output (MIMO)} flows can benefit from the solutions presented in the previous sections. The generic types of MIMO flows are described in \cite{VKTS07}, two of which are shown in Figure \ref{fig:MIMOexamples}. A main difference between SISO and MIMO flows is that apart from re-ordering tasks, additional optimization operations can apply. As explain in \cite{671SVS05}, the \emph{factorize} and \emph{distribute} operations can move an activity appearing in both input subflows of a binary activity to its output and the other way around, respectively.\footnote{\cite{671SVS05} additionally considers the case that an activity can be further split in several sub-activities, which is not considered here.} This allows for example a filtering operation initially placed after a merge task to be pushed down to the merge inputs (provided that the filtering condition refers to both inputs), which is
known to yield better performance.

As we can see in Figure \ref{fig:MIMOexamples}, the \emph{MIMO} flows consist of sub-linear flows. Therefore, the optimization of \emph{SISO} data flows can play an important role in optimizing \emph{MIMO} flows. Algorithm \ref{alg:MIMO} describes a proposal for optimizing \emph{MIMO} flows, which is based on the extraction of the linear segments of the flow and apply optimization algorithms only on the \emph{SISO} sub-flows. Then, we check whether we can apply the factorize/distribute operations, which modify the linear segments. This process is repeated until it converges. In this work, we focus solely on task re-ordering (which corresponds to optimize the linear segments individually) and the investigation of further techniques that combine task re-orderings with additional operations is left for future work.



\section{Experimental Analysis}
\label{experiments}

In this section we present a set of experiments, which have been conducted in order to evaluate the following two factors:
\begin{itemize}
\item {\it Performance} optimization, which corresponds to the minimization of the estimated flow execution cost \emph{SCM}. The performance improvements are measured as the percentage of the decrease in \emph{SCM} after optimization.
\item {\it Time Overhead}, in terms of real time that the generation of the optimized execution plan requires.
\end{itemize}

We construct synthetic flows so that we thoroughly evaluate the algorithms in a wide range of parameter combinations, so that we can derive unbiased and generically applicable lessons for the behaviour of each algorithm.
The main configurable parameters are three:
(i) the number of tasks $n$ ranging from  $10$ up to $100$ (without including the source and the sink tasks) thus covering a range from small to very large data flows;
(ii) the cost and selectivity values of the flow tasks, which are distributed in the range of $[1,100]$ and $(0,2]$, respectively (following either the uniform or the beta distribution); and
(iii) the number of precedence constraints between the flow tasks; in general we consider cases where there are $\alpha\frac{n(n-1)}{2}$ constraints, where $\alpha \in [0.1, 0.98]$. The larger the $\alpha$ value, the less the opportunities for optimization exist. For small $\alpha$ values, there are few PCs, which implies the existence of several valid re-orderings. However, when $\alpha$ becomes 0, the problem reduces to filter ordering in database queries without precedence constraints and thus is out of our interest. Remember that in real cases, we expect PCs to be above 30\%.

In order to conduct the experiments, we randomly generate PC DAGs and task characteristics in a simulation environment. Unless otherwise mentioned, every experiment is repeated 100 times and the average values are presented.
When discussing real times, we use a machine with an Intel Core i5 660 CPU and 6 GB of RAM.

\subsection{Performance Improvements}
In the beginning of Section \ref{sec:ro}, we presented the significant gap between the best performing heuristics to date, namely \emph{Swap}, and the accurate solutions for small flows. We extrapolate that this gap remains, if not widens, for larger flows. The main purpose of this part is to show how the rank ordering-based solutions are capable of filling this gap, and then we discuss the performance benefits due to parallelism in SISO flows. Finally, we evaluate the proposals for MIMO flows.

%
%
 algorithms is presented. As we can observe, there are cases where the \emph{topSort} algorithm has 74\% better performance improvement than the best heuristic. These findings highlighted the need for proposing new approximate optimization methodologies, in order to provide more near-optimal flow execution plans.

\subsubsection{Performance of Rank Ordering-based Solutions}

\begin{figure}[tb!]
\centering
\begin{minipage}{0.45\textwidth}
\includegraphics[width=1.1\textwidth]{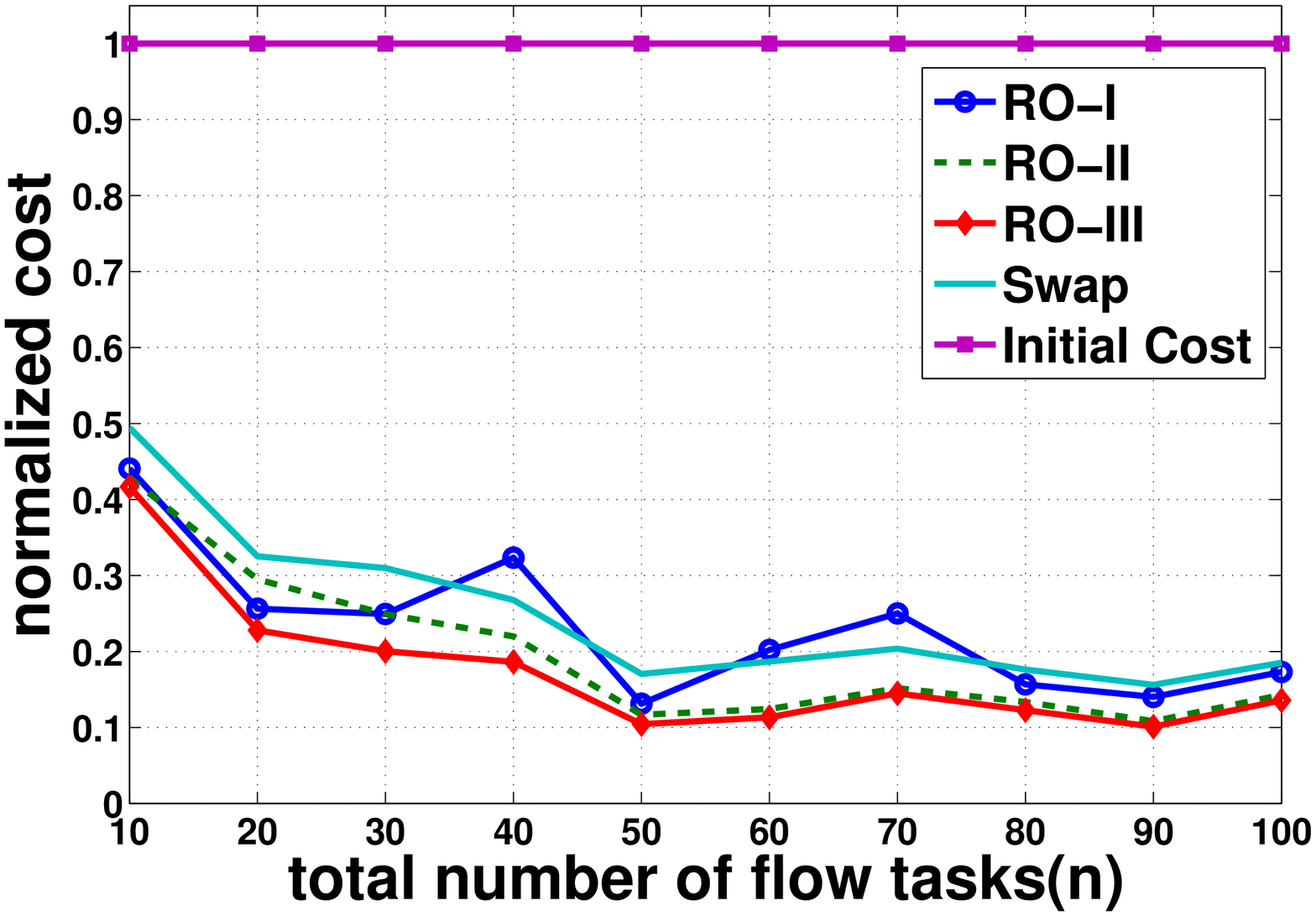}
\end{minipage}
\begin{minipage}{0.45\textwidth}
\includegraphics[width=1.1\textwidth]{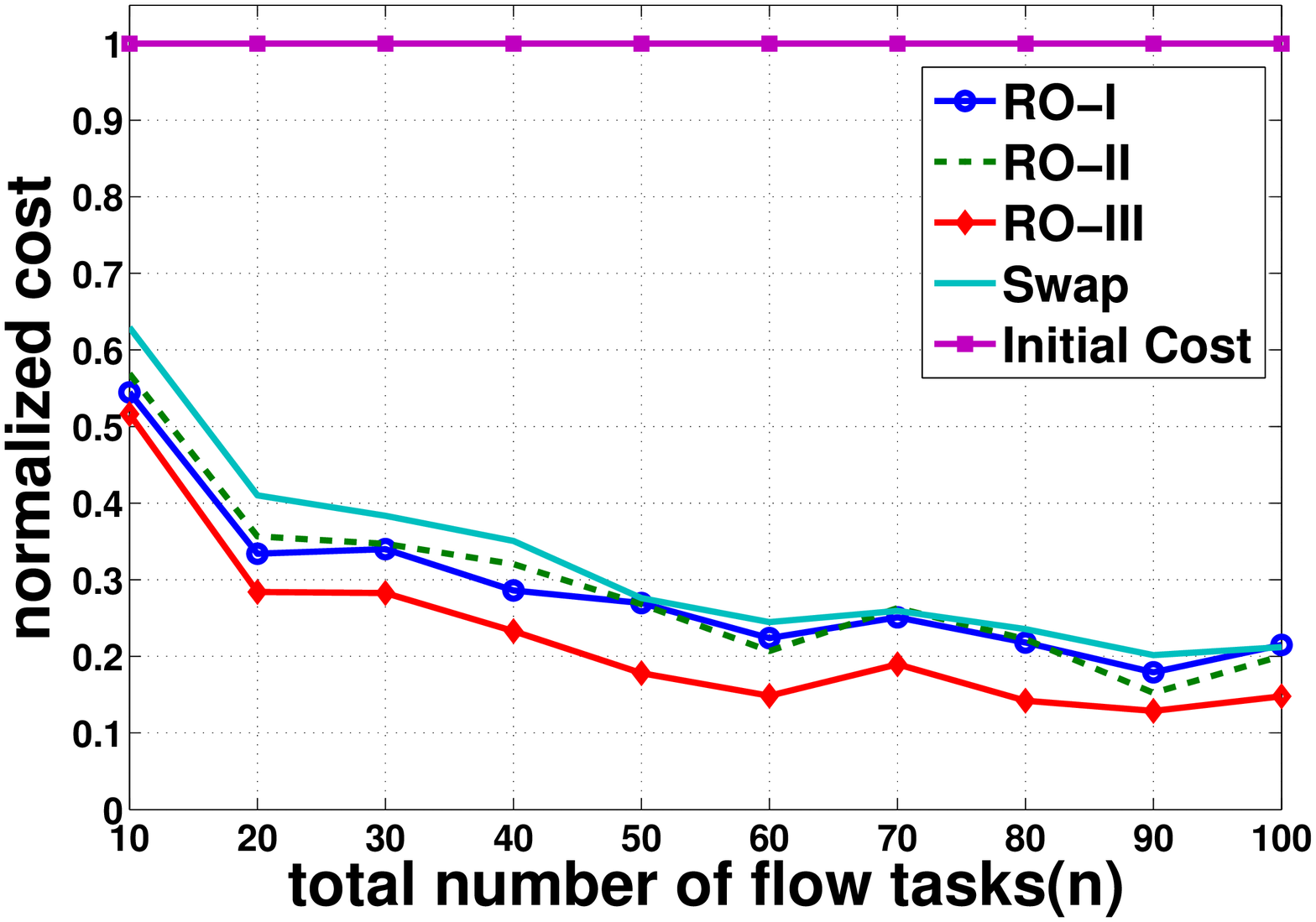}
\end{minipage}
\begin{minipage}{0.45\textwidth}
\includegraphics[width=1.1\textwidth]{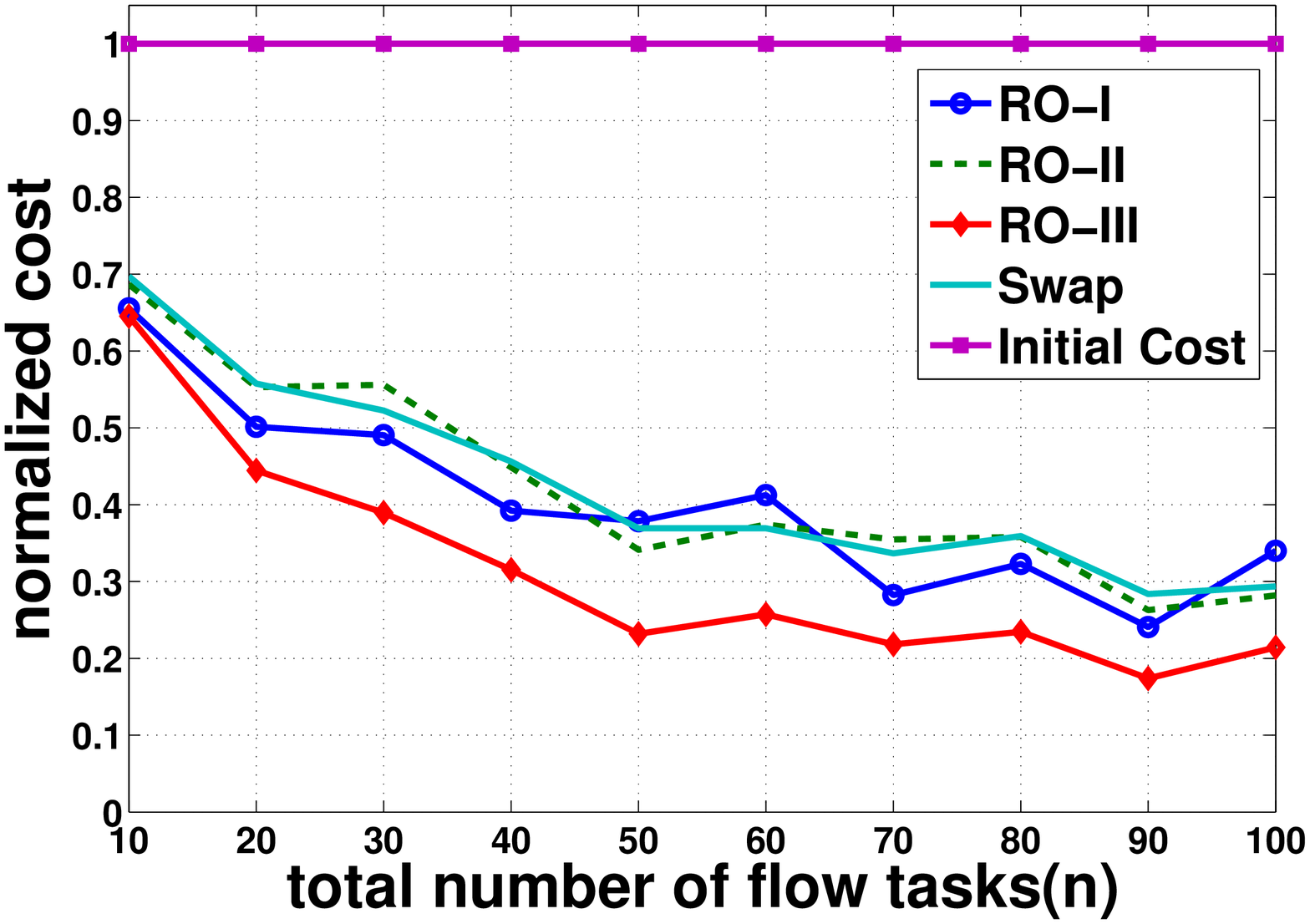}
\end{minipage}
\begin{minipage}{0.45\textwidth}
\includegraphics[width=1.1\textwidth]{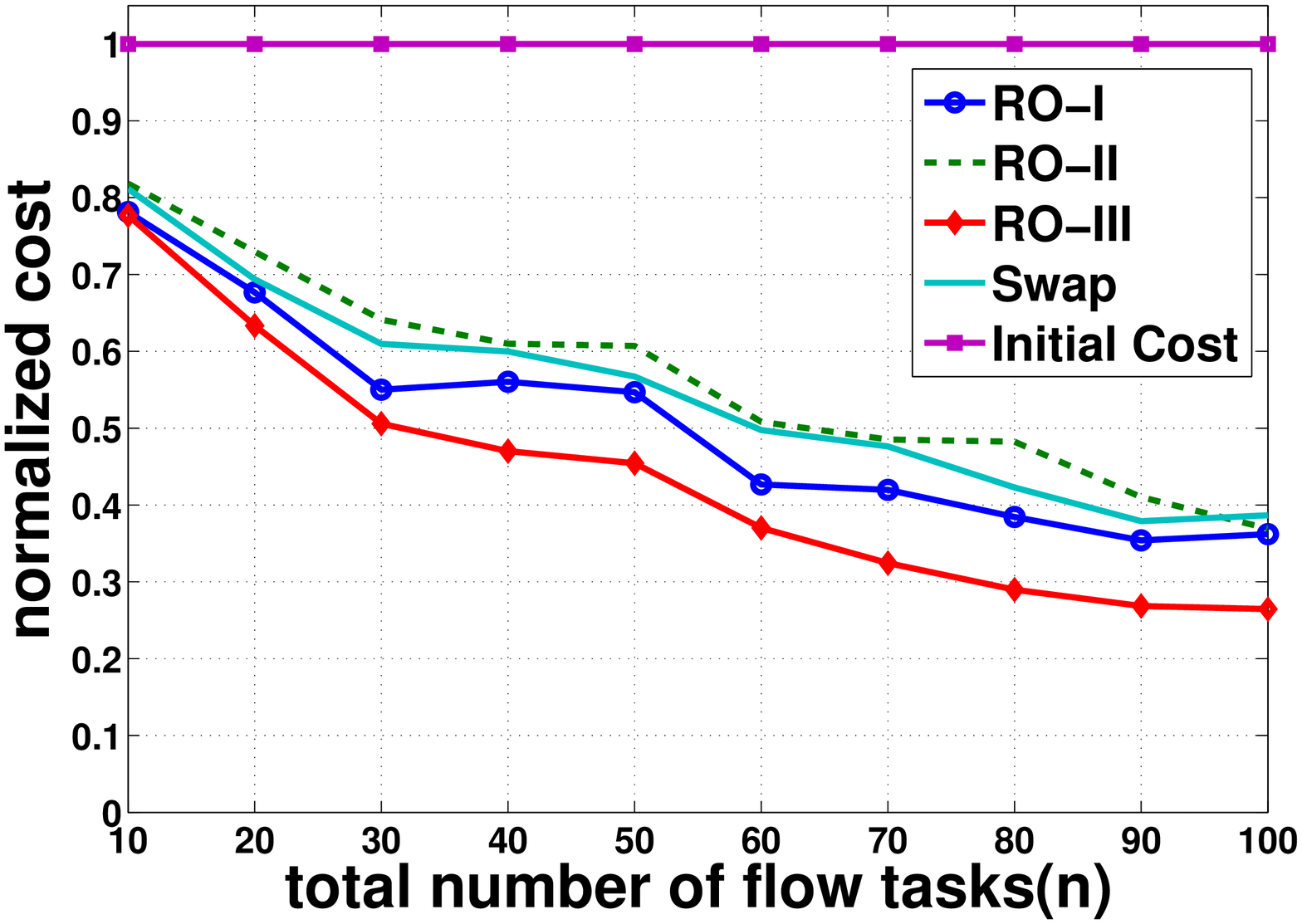}
\end{minipage}
\caption{Improvements in the SCM metric for PCs=20\% (top-left), for PCs=40\% (top-right), for PCs=60\% (bottom-left)and for PCs=80\% (bottom-right).}
\label{fig:ROs}
\vspace{-0.3cm}
\end{figure}

Figure \ref{fig:ROs} presents the results of the comparison of rank ordering-based optimization methodologies with the initial flow execution plan and \emph{Swap}.
The values of the results are normalized according to the performance of the initial (random) execution plan. The four sub-figures present the performance improvement of each optimization proposal for $PCs=20\%,40\%,60\%,80\%$, respectively. Based on these results, a main observation is that \emph{RO-III} is a clear winner, as it outperforms all the other optimization algorithms on average for all the PC percentages examined.  The lesson is that the average improvements of \emph{RO-III} over \emph{Swap} can be significant, as the \emph{RO-III} can yield up to 41\% better performance than \emph{Swap} on average; this difference is observed  for $n=80$ and PC=$40\%$, and means that \emph{RO-III} is on average 1.69 times faster than \emph{Swap} for that case. In addition, the maximum observed speed-up in isolated cases is much higher. For example, in one run where $n=60$ and PC=$60\%$, we have observed a speed-up of more than 73 times in favor of \emph{RO-III}. In another run for $n=100$ and PC=$40\%$, the
speed-up exceeded 285 times (two orders of magnitude).

\emph{RO-I} seems to outperform \emph{RO-II} for 80\% precedence constraints on average, however, if we zoom on the isolated runs, in a significant portion of plans, \emph{RO-II} is better.
For less precedence constraints, there is not a clear winner between \emph{RO-I} and \emph{RO-II}.

Another significant observation from this figure, combined with Figure \ref{fig:heuristic}, is that \emph{RO-III} eliminates the gap between approximate and accurate solutions for 15-task flows. This provide strong insights into the near-optimality of  \emph{RO-III} in practice although no real experiments are feasible in order to establish the ground truth for bigger flows and near optimality cannot be theoretically proved (most probably), as explained in Section \ref{sec:background}.



The experiments above refer to uniformly distributed values of costs and selectivities. We repeat the experiments, when those values follow the beta distribution, which can describe selectivities, as explained in \cite{BC05}.
We have tested several parameters of that distribution, without big differences; here we present the results when the two main beta distribution parameters are set to $a=b=0.5$.
Table \ref{tab:betaTable} presents the results of performance improvement of the \emph{RO-I}, \emph{RO-II}, \emph{RO-III} and \emph{Swap} heuristics normalized according to the cost of the initial randomly generated plan; the PCs are 40\%. The last two columns of Table \ref{tab:betaTable} are computed as follows over all 100 iterations:
\emph{AvgDiff}$=\frac{1}{100}\sum\frac{Swap - ROIII}{Swap}$ and \emph{MaxDiff}$= max\{\frac{Swap - ROIII}{Swap}\}$, and as such the closer the values to 1 the bigger the relative improvement of \emph{RO-III}.

The main observation here is that for beta-distributed values, the performance of \emph{RO-III} against \emph{Swap} improves even more. In the case of flows that consist of 80 and 100 tasks, the \emph{RO-III} results in 60\% and 57\% less \emph{SCM}, which implies a 2.5x and 2.32x speed-up, respectively; this reduction is significantly higher than the one for uniformly distributed metadata. Interestingly, in a specific iteration, the maximum observed decrease is by 3 orders of magnitude. In general, especially for large flows, the performance improvements for beta-distributed values are higher for all techniques.

\begin{table}[tb!]
\centering \setlength\tabcolsep{5pt}
\caption{Normalized performance for data flows with 40\% engine constraints}
\label{tab:betaTable}
\scriptsize
\begin{tabular}{|c||c|c|c|c|c||c|c|}
\hline
\multicolumn{8}{|c|}{\textbf{Uniformly Distributed Cost and Selectivity Values}}\tabularnewline\hline
$\textbf{\textit{n}}$ & $\textbf{\textit{Initial}}$ & $\textbf{\textit{RO-I}}$ & $\textbf{\textit{RO-II}}$ & $\textbf{\textit{RO-III}}$ & $\textbf{\textit{Swap}}$ & $\textbf{\textit{Avg Diff}}$ & $\textbf{\textit{Max Diff}}$\\ \hline \hline
$\textbf{\textit{20}}$ & 1.0000 & 0.3339 & 0.3566 & 0.2841 & 0.4101 & 0.2636 & 0.8102 \\ \hline
$\textbf{\textit{50}}$ & 1.0000 & 0.2696 & 0.2679 & 0.1780 & 0.2761 & 0.3281 & 0.9802 \\ \hline
$\textbf{\textit{80}}$ & 1.0000 & 0.2181 & 0.2225 & 0.1420 & 0.2355 & 0.4069 & 0.9663 \\ \hline
$\textbf{\textit{100}}$ & 1.0000 & 0.2149 & 0.2005 & 0.1478 & 0.2120 & 0.2900 & 0.9965 \\ \hline
\multicolumn{8}{|c|}{\textbf{Beta Distributed Cost and Selectivity Values}}\tabularnewline\hline
$\textbf{\textit{n}}$ & $\textbf{\textit{Initial}}$ & $\textbf{\textit{RO-I}}$ & $\textbf{\textit{RO-II}}$ & $\textbf{\textit{RO-III}}$ & $\textbf{\textit{Swap}}$ & $\textbf{\textit{Avg Diff}}$ & $\textbf{\textit{Max Diff}}$\\ \hline \hline
$\textbf{\textit{20}}$ & 1.0000 & 0.3509 & 0.4235 & 0.2837 & 0.4035 & 0.2756 & 0.9562 \\ \hline
$\textbf{\textit{50}}$ & 1.0000 & 0.3942 & 0.2287 & 0.1075 & 0.2310 & 0.4865 & 0.9898 \\ \hline
$\textbf{\textit{80}}$ & 1.0000 & 0.1356 & 0.1945 & 0.0553 & 0.1403 & 0.6041 & 0.9949 \\ \hline
$\textbf{\textit{100}}$ & 1.0000 & 0.0944 & 0.1591 & 0.0538 & 0.1141 & 0.5699 & 0.9995 \\ \hline
\end{tabular}
\end{table}

\subsubsection{Performance of Parallel Optimization Solutions}

\begin{table}[tb!]
\centering \setlength\tabcolsep{5pt}
\caption{Normalized performance for data flows with n=50,100 tasks.}
\label{tab:extraCostTable}
\footnotesize
\begin{tabular}{|r||c|c|c|c|}
\hline
\multicolumn{5}{|c|}{\textbf{n=50}} \tabularnewline\hline \hline
$\textbf{alg$\backslash$PCs(\%)}$ & $\textbf{\textit{20}}$ & $\textbf{\textit{40}}$ & $\textbf{\textit{60}}$ & $\textbf{\textit{80}}$ \\ \hline
$\textbf{Initial}$ & 1.0000 & 1.0000 & 1.0000 & 1.0000 \\ \hline
$\textbf{PSwap}$ & 0.1759 & 0.2723 & 0.3329 & 0.4987 \\ \hline
$\textbf{PSwap$'$}$ & 0.1804 & 0.2812 & 0.3448 & 0.5177 \\ \hline
$\textbf{PGreedyII}$ & 0.1052 & 0.1839 & 0.2842 & 0.4413 \\ \hline
$\textbf{PGreedyII$'$}$ & 0.1057 & 0.1865 & 0.2921 & 0.4552 \\ \hline
$\textbf{PRO-I}$ & 0.1340 & 0.2277 & 0.2949 & 0.4534 \\ \hline
$\textbf{PRO-I$'$}$ & 0.1363 & 0.2321 & 0.3011 & 0.4629 \\ \hline
$\textbf{PRO-II}$ & 0.1171 & 0.2418 & 0.4455 & 0.5355 \\ \hline
$\textbf{PRO-II$'$}$ & 0.1188 & 0.2497 & 0.4686 & 0.5579 \\ \hline
$\textbf{PRO-III}$ & 0.0989 & 0.1600 & 0.2156 & 0.4012 \\ \hline
$\textbf{PRO-III$'$}$ & 0.0990 & 0.1605 & 0.2166 & 0.4062 \\ \hline
\hline

\multicolumn{5}{|c|}{\textbf{n=100}} \tabularnewline\hline \hline
$\textbf{alg$\backslash$PCs(\%)}$ & $\textbf{\textit{20}}$ & $\textbf{\textit{40}}$ & $\textbf{\textit{60}}$ & $\textbf{\textit{80}}$ \\ \hline
$\textbf{Initial}$ & 1.0000 & 1.0000 & 1.0000 & 1.0000 \\ \hline
$\textbf{PSwap}$ & 0.0855 & 0.1428 & 0.2087 & 0.3440 \\ \hline
$\textbf{PSwap$'$}$ & 0.0886 & 0.1488 & 0.2197 & 0.3580 \\ \hline
$\textbf{PGreedyII}$ & 0.0485 & 0.0765 & 0.1274 & 0.2635 \\ \hline
$\textbf{PGreedyII$'$}$ & 0.0485 & 0.0769 & 0.1299 & 0.2719 \\ \hline
$\textbf{PRO-I}$ & 0.0793 & 0.1264 & 0.2013 & 0.2994 \\ \hline
$\textbf{PRO-I$'$}$ & 0.0820 & 0.1302 & 0.2072 & 0.3069 \\ \hline
$\textbf{PRO-II}$ & 0.0605 & 0.4507 & 0.2522 & 0.4073 \\ \hline
$\textbf{PRO-II$'$}$ & 0.0618 & 0.4911 & 0.2671 & 0.4278 \\ \hline
$\textbf{PRO-III}$ & 0.0465 & 0.0681 & 0.1058 & 0.2183 \\ \hline
$\textbf{PRO-III$'$}$ & 0.0465 & 0.0681 & 0.1063 & 0.2204 \\ \hline
\end{tabular}
\end{table}


This set of experiments is conducted in order to evaluate the performance of data flows when they are executed in parallel according to the techniques discussed in Section \ref{sec:parallel}. To this end, we compare   the parallel version of \emph{Swap}, named as \emph{PSwap}, against the parallel proposed rank ordering-based algorithms, denoted as \emph{PRO-I,PRO-II,PRO-III}, respectively.  We also compare against \emph{PGreedyII}, which outperforms \emph{PGreedyI} as shown in additional experiments in \ref{sec:app-pgreedy}. Initially, we assume that the merge cost $mc$ is 0, but we relax this assumption later.

The comparisons are presented in Table \ref{tab:extraCostTable}, where it is shown that the parallelized version of \emph{RO-III}, \emph{PRO-III}, strengthens its position as the best performing technique.
When the merge cost is considered, the names of the algorithms are coupled with the prime symbol; for the moment we do not focus on those table rows.
For linear flows, when $n$=50 and PCs=40\%, RO-III results in
decrease of the \emph{SCM} of \emph{Swap} by 32\% (see Table \ref{tab:betaTable}). In a parallel setting, the decrease in SCM comparing \emph{PSwap} and \emph{PRO-III} reaches 41\%. Also, for n=100, the performance improvements reach 52\% (from 29\%, see Table \ref{tab:betaTable}). The relative improvements are similar for PCs=60\% and slightly less for PCs = 20\% and PCs= 80\%.

A question arises as to how often parallelization leads to benefits. Analyzing the individual runs, we have observed that the number of such occurrences is less than 10\%, if we count only improvements higher than 2\%. Nevertheless, the magnitude of the improvements is strong;y correlated with the number of PCs. For less constraints settings (PCs = 20\%), for both $n$=50 and $n$=100, we have observed speed-ups of an order of magnitude. When PCs=40\%, the maximum observed speed-up drops to 4 and 3 times, respectively. For even more PCs, this speed-up does not exceed 12.7\%. A final note is that \emph{PGreedyI} is the best performing parallel heuristic from those not fully proposed in this work. The main conclusion up to here is that further refining the linear orderings with our proposed light-weight post-processing step can yield tangible performance improvements, and our proposals lead to further advancements in the current state-of-the-art in linear flow optimization.


Next, we repeat the experiments with non-zero merge cost, and the results verify that its impact is negligible (see Table \ref{tab:extraCostTable}). After real experiments with the PDI tool, we set $mc=10$, that is an order of magnitude higher than the less expensive tasks and an order of magnitude lower than the most expensive ones.
Overall, on average, our best performing solution, namely \emph{PRO-III} continues to have average performance improvements against \emph{Swap} of an order of magnitude.


\subsubsection{Performance of MIMO flows}
\begin{figure}[tb!]
\centering
\includegraphics[width=0.50\textwidth]{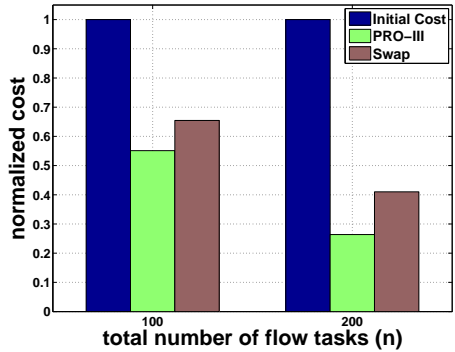}
\caption{MIMO optimization for n=100,200 and 40\% precedence constraints}
\label{fig:mimoExp}
\end{figure}

This set of experiments considers the evaluation of the methodology that is analyzed in Section \ref{sec:mimo} for \emph{\emph{MIMO}} data flow optimization. We consider two cases of butterfly flows (see Figure \ref{fig:MIMOexamples}(left)). In each case we consider 10 linear segments with 10 and 20 tasks, respectively; thus the overall number of tasks is 100 and 200. The percentage of PCs is 40\%.

Figure \ref{fig:mimoExp} presents the average performance improvements of the \emph{PRO-III} and \emph{Swap} algorithms over the non-optimized initial data flow. In the case where the linear segments are very small (10 tasks) the improvements are small as well. When the linear segment size increases to 20, \emph{PRO-III} has 34\% better performance improvement than \emph{Swap}, and 74\% lower execution cost compared to the non-optimized case.
The performance improvements are commensurate with those in Table \ref{tab:betaTable}, which supports are claim that our proposals for SISO flows can be transferred to MIMO settings as well.

\subsection{Time Overhead}

\begin{figure}[tb!]
\centering
\begin{minipage}{0.45\textwidth}
\includegraphics[width=1.1\textwidth]{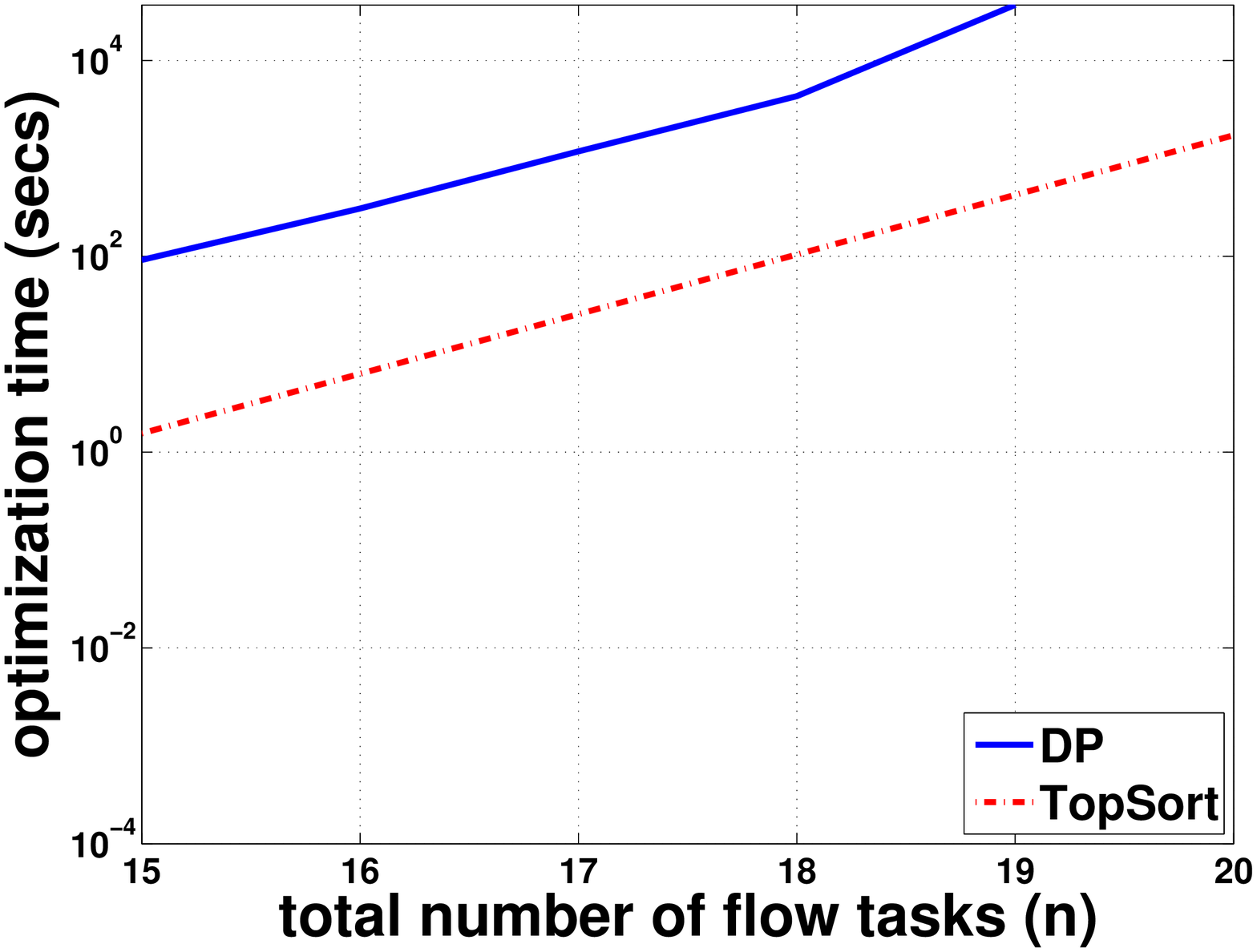}
\end{minipage}
\begin{minipage}{0.45\textwidth}
\includegraphics[width=1.1\textwidth]{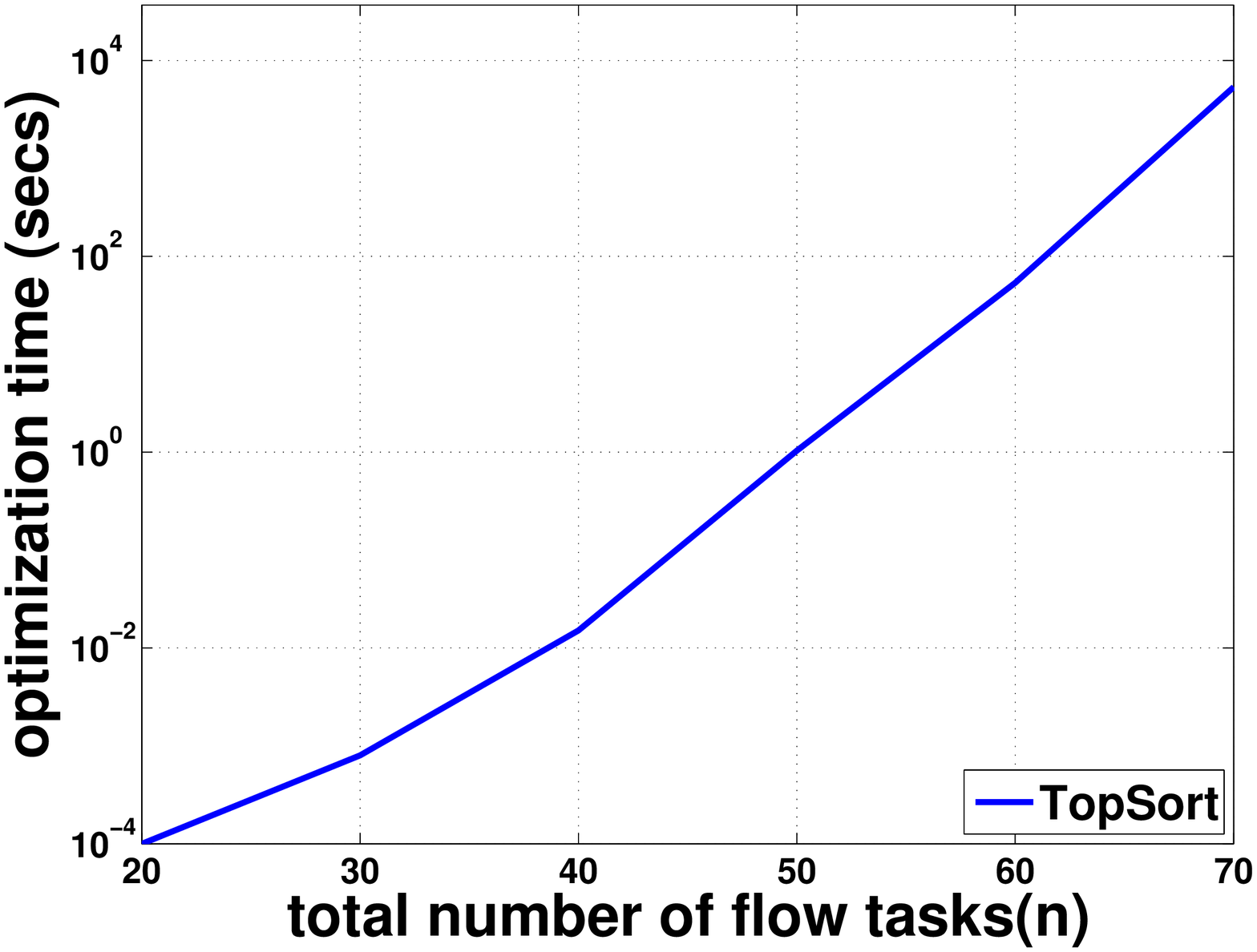}
\end{minipage}
\begin{minipage}{0.45\textwidth}
\includegraphics[width=1.1\textwidth]{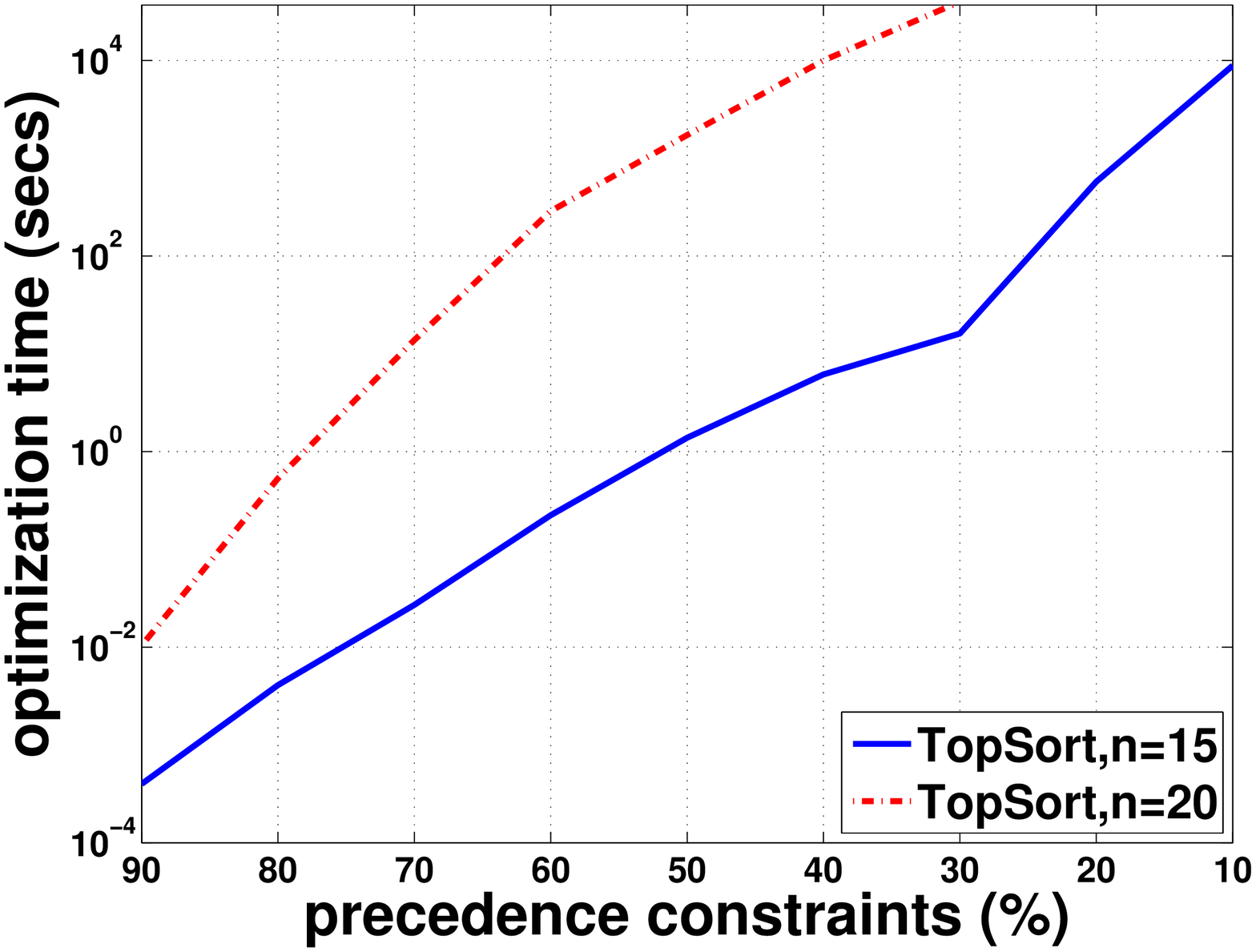}
\end{minipage}
\begin{minipage}{0.45\textwidth}
\includegraphics[width=1.1\textwidth]{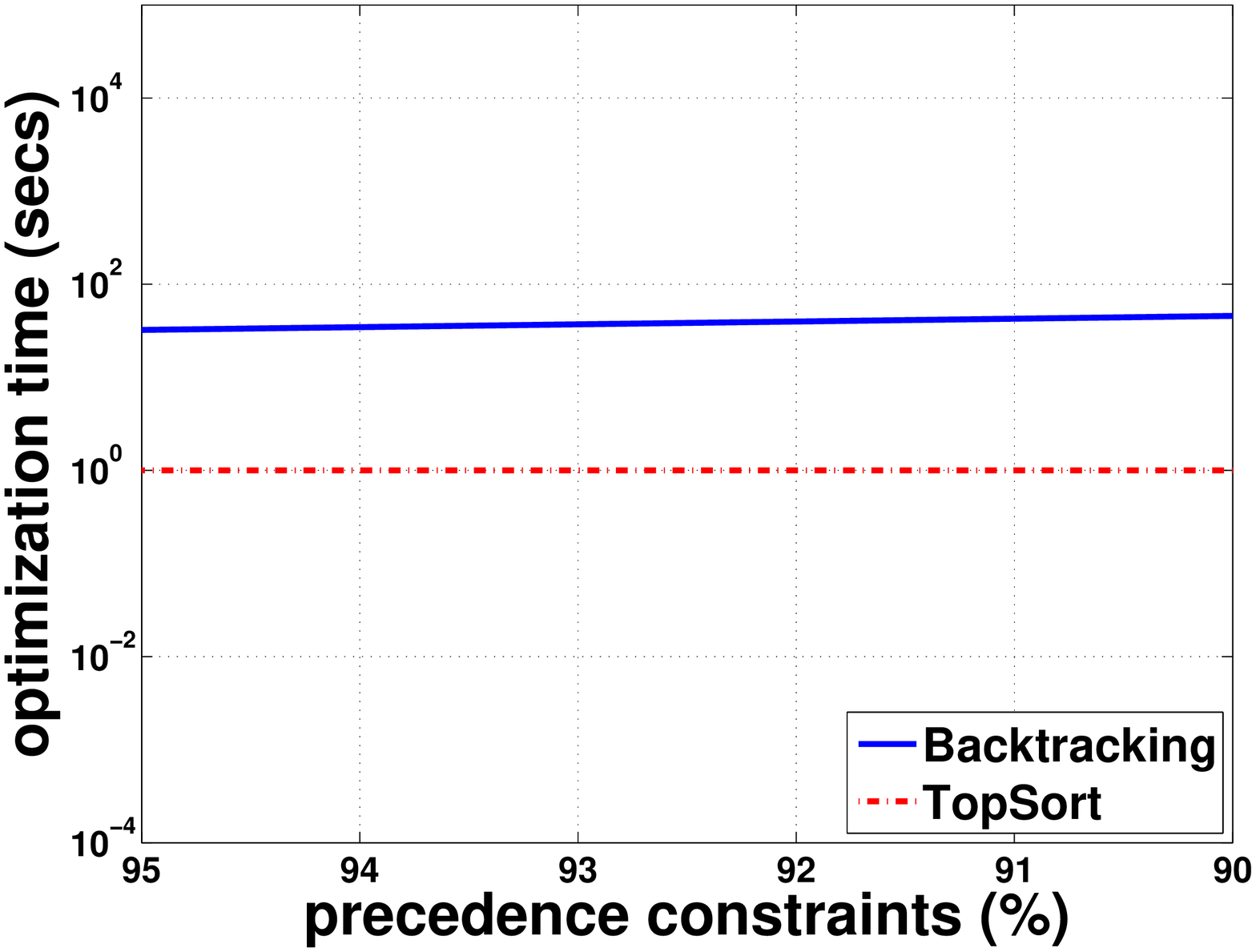}
\end{minipage}
\caption{Optimization overhead for \emph{DP} and \emph{topSort} with 50\% precedence constraints and n = 15,...,20 (top-left), for \emph{TopSort} when n = 10,...,70 and PCs=98\% (top-right), and n = 15,20 and PCs vary (bottom-left), and for \emph{Backtracking} and \emph{TopSort} when n = 15 with different range of precedence constraints (bottom-right).}
\label{fig:exhaust}
\vspace{-0.3cm}
\end{figure}

In this section, we conduct a thorough evaluation of the time overhead of the accurate optimization algorithms. The purpose of this set of experiments is to show that the application of the exhaustive algorithms, and more specifically of \emph{TopSort},  is limited only to small or very constrained flows. Figure \ref{fig:exhaust}(top-left) presents the average execution time of the \emph{DP} algorithm compared to the \emph{TopSort} solution for 50\% precedence constraints. More specifically, this figure depicts the time overhead for executing data flows with $n=15,...,20$ flow tasks. The main conclusions that can be drawn from this figure is that \emph{DP} algorithm is not a practical optimization solution  even for small flows that consist of 19 flow activities; the execution of a flow with 20 tasks requires over 3 days using our test machine. Even if the \emph{TopSort} algorithm runs at least 50 times faster than \emph{DP}, the execution of \emph{TopSort} follows a similar pattern with \emph{DP}.

Figure \ref{fig:exhaust}(top-right) shows the average execution time of \emph{TopSort} for flows with $n=10,...,70$ having 98\% precedence constraints, which implies that the number of the possible re-orderings is quite restricted. \emph{TopSort} does not scale well, but can run in acceptable time even for medium-sized flows of 60 tasks.
Additionally, Figure \ref{fig:exhaust} (bottom-left) depicts that \emph{TopSort} cannot scale for arbitrary precedence constraints even for flows with 15 and 20 flow activities. For example, the execution time of a data flow with 20 tasks and 50\% precedence constraints is 2 orders of magnitude higher than the execution time of a data flow with 15 tasks.
Finally, in the bottom-right part of Figure \ref{fig:exhaust}, the time overhead of \emph{Backtracking} compared to \emph{TopSort} is presented. The main observation of this figure, where the precedence constraints range is $PCs=90\%,...,98\%$, is that  \emph{Backtracking} can be up to 62 times slower than \emph{TopSort}.

Overall, we can conclude that \emph{TopSort}, on the one hand scales better than the other techniques and is applicable in specific cases where the other two approaches are not, but, on the other, it is not able to scale in general.
We do not present the overhead of the approximate solutions, because it is negligible.

\section{Related Work}
\label{rw}

The existing approaches of flow optimization can be classified in the following main categories, which are subsequently presented in turn:
\begin{itemize}
\item {\it Optimization of the structure of data flows:} this category targets the methodologies that optimize the flow execution plan through changes in the structure of the flow graph including task re-ordering.
\item {\it Optimization of the resource allocation and scheduling aspects of data flows:} the proposals in this category deal with issues such as the allocation of computational resources and specific execution engines to each part of the flow along with time scheduling details, without affecting the workflow structure.
\item {\it Application-dependent solutions:} this category contains optimization techniques that are specific to certain settings; interestingly, some of these techniques leverage database technologies.
\end{itemize}

{\it Optimization of the structure of data flows.}
An  aspect of this category that is particularly relevant to our work considers flow optimization inspired by query processing techniques.  In \cite{Flor99}, an optimization algorithm for query plans with dependency constraints between algebraic operators is presented. The adaptation of this algorithm in our SISO problem setting that does not consider only algebraic operators is reduced to the existing optimization algorithms we have presented in previous sections, and more specifically to {\it GreedyI} and {\it Partition}. In \cite{KG12}, ad-hoc query optimization methodologies are employed in order to perform structure reformations, such as reordering and introducing new services in an existing workflow; in this work we investigate more systematic approaches.

Optimizations of Extract Transform Loading (ETL) flows are analyzed in \cite{671SVS05}. Specifically, the authors consider ETL execution plans as states and use transitions, such as swap, merge, split, distribute and so on, to generate new states in order to navigate through the state space, which corresponds to the execution plan alternatives; they also present optimization algorithms for reducing ETL workflow execution cost albeit with exponential complexity. In our work, where we consider only task re-orderings, the proposal in  \cite{671SVS05} corresponds to the \emph{Swap} algorithm, which we have presented and evaluated.

Another interesting approach to flow optimization is presented in \cite{HPS+12}, where the optimizations are based on the analysis of the properties of user-defined functions that implement the data processing logic. This work focuses mostly on techniques that infer the dependency constraints between tasks through examination of their internal semantics rather than on task reordering algorithms per se.   In \cite{DatInt09},  they introduce a suite of quality metrics (QoX) without going into flow optimization algorithm details.

In addition, there is a significant portion of proposals on flow optimization that proceed to flow structure optimizations but do not perform task reordering, as we do. For example, an interesting proposal that aims to combine the control and the data flow view of workflows  has appeared in \cite{638VSS+07}. That work presents approaches that merge tasks related to data management to decrease the number of invocations to the underlying databases without changing the relative order of the tasks. In \cite{MWS00}, a data oriented method for workflow optimization is proposed in order to minimize execution cost. This method is based on the fact that data may be shared across several functions, and, as such, workflow performance stands to benefit from optimizations in the form of incorporating a shared database to handle common data-oriented tasks. Another workflow optimization method that affects the workflow structure with a view to improving the efficiency of the workflow is presented in \cite{DSW98}. This
method is inspired by the current limitations of business information processes. In particular, a task redesigning method is presented, which is based on the consolidation of the tasks to reduce the overall execution time. Quality of Service requirements (QoS), such as precedence of information flows and technology support costs are taken into account.
In \cite{Tziov07}, a methodology to choose the optimal physical implementation of each task and decide whether to introduce special sorting tasks is presented, when there are several implementation alternatives. This work does not consider the execution order of the flow activities. Several optimizations in workflows are also discussed in \cite{Bohm2011}, but the techniques are limited to straightforward application of query optimization techniques, such as join reordering and pushing down selections.

{\it Optimization of the resource allocation and scheduling aspects of data flows.}
The main motivation of the proposals in this category stem from the need for more efficient resource management, given that resource management is deemed as a key performance factor. Contrary to our work, they assume an execution setting with multiple execution engines and do not deal with optimization of the flow task ordering.
For example, in \cite{DPF07,XZHY07}, they introduce resource allocation algorithms and heuristic techniques that have the ability to take into account constraints, such as cost optimization, user-specified deadline and workflow partitioning according to assigned deadlines \cite{XZHY07}.  \cite{ODOPVM11} discusses methodologies about how to execute and dispatch task activities in parallel computers.

Another family of optimization proposals deals with task scheduling methods, considering aspects such as semantic expression of workflow tasks, dynamic selection of services among many candidates and latency minimization \cite{SWH08,ChenZ09,KMRHK09,ABMR10}. Also, there are scheduling methods which are exclusively related with grid workflow optimization (e.g., \cite{SWH08,ChenZ09,KMRHK09}), or linear workflow optimization, such as \cite{ABMR10}, which discusses optimal time schedules given a fixed allocation of activities to engines. Also, a set of optimization algorithms based on deadline and time constraints was analyzed for scheduling flows in \cite{ANE12,ANE13}. Another proposal of flow optimization is presented in \cite{PP2012} based on soft deadline rescheduling in order to deal with the problem of  fault tolerance in flow executions. In \cite{CB2013}, an optimization methodology for minimizing the performance fluctuations that might occur by the resource diversity, which also considers deadlines, is
proposed. Additionally, there is s set of optimization methodologies based on multi-objective optimization. For example, an auction-based scheduling methodology for multi-objective flow optimization is presented in \cite{FPF13}, while \cite{SWCD12,SWDH13} propose optimization methodologies for multi-engine environments meeting multiple objectives, such as fault-tolerance and performance. The implementation of some of the presented optimization methods mentioned above is carried out with the help of algorithms that take into consideration certain quality of service requirements (QoS). In this case, users are responsible to set constraints, such as reliability, time, security, cost and fidelity, which are the principle parameters of workflow task scheduling. In this work, we do not consider resource allocation and scheduling issues, which are orthogonal to task ordering.

{\it Application-dependent solutions.}
An important part of workflow optimization research was originated by optimization methods that have been created for a specific applications and as such, they are application dependent. An example of application dependent workflow optimization is discussed in \cite{DS02}, which deals with the creation and process of technical documents by a document workflow management system; in this work, the parallelism opportunities presented by the document structure are exploited to optimize workflows. Another example is \cite{BCLSTW05}, where a process execution management framework is proposed in order to optimize business objectives of processes in a dynamic business environment. Also, there are workflow optimization methodologies applied in other scientific fields. A representative example is \cite{HKVR09}, where the optimization algorithms are used for the  development of molecular models and they are applied to a simulation tool. Analogous examples that achieve workflow optimization only under certain
circumstances are presented in \cite{RSC02,FSPT07,HAAWRH07,PLKHK09}. However, these optimization methods cannot be adapted to a more general case.

\section{Conclusions}
\label{conclusions}
In this work, we deal with the problem of specifying the optimal execution order of constituent tasks of a data flow in order to minimize the sum of the task execution costs. We are motivated by the significant limitations of fully-automated optimization solutions for data flows, as, nowadays, the optimization of the complex data flows is left to the flow designers and is a manual procedure. Firstly, as the query optimization techniques are not applicable to data flow optimization because of the precedence constraints and the existing proposals for optimal solutions cannot scale, there is significant need to propose new flow optimization methodologies. We show that the state-of-the-art optimization algorithms can have 74\%  higher execution cost than the optimal solution even for the simplest type of \emph{single-input single-output (SISO)} flows with a small number of tasks. So, to fill the gap of near-optimal optimization techniques, we propose a set of approximate algorithms that can exhibit 40\%
performance improvements than the best existing heuristic. We also introduce a post-process optimization phase for parallel execution of the flow tasks in order to improve even more the performance of a data flow, and we show that we can extend these solutions to more complex data flow scenarios that deal with arbitrary number of sources and sinks. This work aims to provide the basis for more holistic flow optimization algorithms, which do not only consider more complex flows, but also combine task ordering with aspects, such as task implementation and scheduling.

\section{Acknowledgments}
This research has been co-financed by the European Union (European Social Fund - ESF) and Greek national funds through the Operational Program ``Education and Lifelong Learning" of the National Strategic Reference Framework (NSRF) - Research Funding Program: Thales. Investing in knowledge society through the European Social Fund.

\bibliographystyle{model1-num-names}
\bibliography{taskOrderBIB}







\appendix
\section{Extra material about the \emph{DP} algorithm}
\label{sec:app-dp}

\begin{algorithm}[tb!]
\caption{Dynamic Programming}
\label{alg:dp}
\begin{algorithmic}[1]
\begin{small}
\REQUIRE A set of n tasks, T=\{$t_1$, ..., $t_n$\} \\
         A directed acyclic graph PC with precedence constraints
\ENSURE A directed acyclic graph P representing the optimal plan
\\
\COMMENT{Initialize {\it PartialPlan}, {\it Costs} and {\it Sel} of size $2^n-1$}
\FOR{all $i \in \{2, ...., n\}$ }
\STATE PartialPlan[$2^{i-1}]=t_i$; Costs[$2^{i-1}]=c_i$; Sel[$2^{i-1}]=sel_i$];
\ENDFOR

\FOR {all $s \in \{2, ...., n\}$ }
\STATE $R \leftarrow Subsets(T, s)$ \COMMENT{X is a set with all subsets of T of size s}
\\ \COMMENT{r is a specific subset of size s}
\STATE $tempBest \leftarrow \infty$
\FOR{each $r \in R$}
\FOR {all $i \in \{1, ..., r.length()\}$}
\STATE $tempSet \leftarrow r - r(i)$
\STATE $pos1 \leftarrow findIndex(tempSet)$
\STATE $pos2 \leftarrow findIndex(r(i))$
\IF {sp(i) has all predecessors in $tempSet$}
\STATE $TempPlan \leftarrow tempSet,r(i)$
\STATE $costTempPlan \leftarrow Costs[pos1] +  Sel[pos1]Costs[pos2] $
\IF {$costTempPlan < tempBest$}
\STATE $tempBest  \leftarrow costTempPlan $
\STATE $k \leftarrow pos1+pos2$
\STATE update({\it PartialPlan[k]}, {\it Costs[k]}, {\it Sel[k]})
\ENDIF
\ENDIF
\ENDFOR
\ENDFOR
\ENDFOR
\STATE $P \leftarrow PartialPlan[2^{n}-1]$
\end{small}
\end{algorithmic}
\end{algorithm}

\begin{figure}[tb!]
\centering
\vspace{-10pt}
\includegraphics[width=0.6\textwidth]{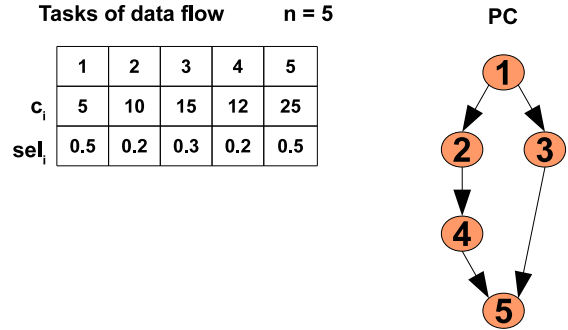}
\caption{Metadata of the example data flow}
\label{fig:algInput}
\end{figure}

\begin{figure}[tb!]
\centering
\vspace{-10pt}
\includegraphics[width=0.75\textwidth]{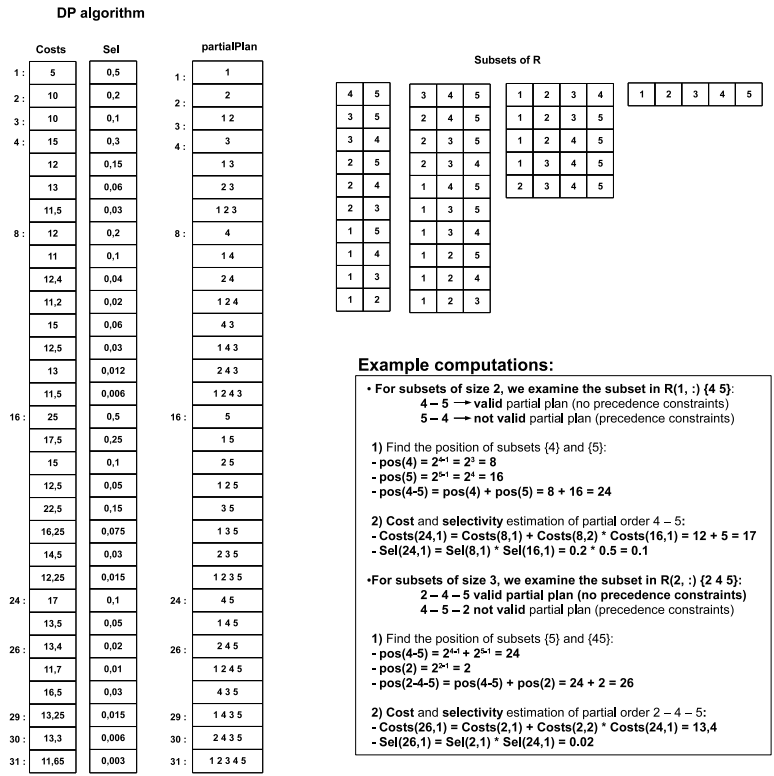}
\caption{Example of the DP algorithm.}
\label{fig:DPexampl}
\end{figure}

In order to implement the algorithm, we use three vectors of size $2^{n}-1$, namely {\it PartialPlan}, {\it Costs} and {\it Sel}. According to the algorithm implementation, the $i$-th cell corresponds to the combination of tasks for which the bit is 1 in its binary representation. For example, if $i=13$, then the binary representation of this position is $(1101)_2$. Specifically, this means that $partialPlan[13]$ corresponds to the optimal ordering of the $1^{st}$, $2^{nd}$ and $4^{th}$ tasks. The {\it Costs} and {\it Sel} vectors hold the aggregate cost and selectivity of the subplans, respectively. The last cell of {\it PartialPlan} and {\it Costs} contain the optimal plan  and its total cost, respectively.
A complete pseudocode is shown in Algorithm \ref{alg:dp}. For the sake of simplicity of presentation, the algorithm is not fully optimized; e.g., in line 18, the update of vertices may occur only once after the final best plan is found.

We give an example of the algorithm with a flow with $n=5$; the task metadata are shown in  Figure \ref{fig:algInput}. The \emph{DP} example is in Figure \ref{fig:DPexampl}.
First of all, all the subsets $R$ of $T$ of length $K=\{1, 2, ..., n\}$ are found. For single task subsets, such as $\{t_1\}, \{t_2\}, ..., \{t_n\}$, \emph{DP }estimates their position in the $partialPlan$ matrix, e.g. $\{2\}$ subset is positioned in $partialPlan(2^{2-1},1)$.  For subsets with length greater than $1$, e.g., the subset $\{1, 3, 4\}$, we examine the case that each element of that subset is placed at the end of the subset. If the precedence constraints are violated, \emph{DP} continues to the next placement. If the precedence constrains are not violated, the algorithm estimates the cost of the valid partial plan with that element positioned at the end of the subset, reusing the results of the orderings of smaller subsets. Similarly, the cost of all orderings in the subset is estimated and the algorithm finds the ordering of the subset with the minimum cost. The optimal partial plan, its cost and the product of task selectivities are stored in the corresponding position in the $partialPLan$ and
$DPcs$ vertices, respectively. For example, the partial plan $\{1, 3, 4, 5\}$ is stored in position $2^{1-1} + 2^{3-1} + 2^{4-1} + 2^{5-1} = 29$  of the $partialPlan$ matrix.

{\it Correctness}: If {\it PartialPlan} is of size $n=1$, the optimal solution is trivial and is found by the algorithm during initialization in lines 1-3 of Algorithm \ref{alg:dp}. We assume that a {\it PartialPlan} of size $n-1$ is optimal and we need to prove that  {\it PartialPlan} of size $n$ is also optimal. The sketch of the proof will be based on contradiction. Let us assume that the \emph{DP} does not produce the optimal solution. Any linear solution of size $n$ consists of a {\it PartialPlan} of size $n-1$ followed by the $n$-th task; \emph{DP} checks all the alternatives for the $n$-th task. So, there is a different optimal solution, where the {\it PartialPlan} of size $n-1$ is different of \emph{DP}'s {\it PartialPlan} of the same size. According to the \emph{SCM}, the cost of the subplan of size $n$ is computed as the sum of two components: the cost of subplan of size $n-1$ and the cost of the $n$-th task times the selectivity of the first $n-1$ tasks. The costs of the solutions of size $n$,
which end with the same task, differ only in the first component. According to our assumptions, the cost of \emph{DP}'s {\it PartialPlan} of size $n-1$ cannot be higher than any other  subplan solution of size $n-1$ by definition. Consequently, there is no other solution different from \emph{DP}'s solution that can yield lower cost. This completes the proof.

\section{Extra material about the \emph{TopSort} algorithm}
\label{sec:app-ts}

\begin{algorithm}[tb!]
\caption{TopSort}
\label{alg:topSort}
\begin{algorithmic}[1]
\begin{small}
\REQUIRE A set of n tasks, T=\{$t_1$, ..., $t_n$\} with known costs and selectivities.\\A directed acyclic graph $PC$ with precedence constraints.\\
\ENSURE An ordering of the tasks P representing the optimal plan.
\STATE G=\{$t_1$, $t_2$, ..., $t_n$\} \COMMENT{G is initialized with a valid topological ordering ordering of $PC$.}
\STATE i=1
\STATE minCost $\leftarrow$ computeSCM($G$)
\WHILE {{$i<n$} \COMMENT{n is the total number of tasks}}
\STATE k $\leftarrow$ location(1,i)
\STATE k1 $\leftarrow$ k + 1
\IF {$G(k1)$ task has prerequisite $i$}
\STATE {$//$ \bf{Rotation stage}}
\STATE Rotate the  elements of G from positions $i$ to $k$
\STATE cost $\leftarrow$ computeSCM($G$)
\STATE i$\leftarrow$ i+1
\ELSE
\STATE {$//$ \bf{Swapping stage}}
\STATE Swap the $k$ and $k1$ elements of G
\STATE cost $\leftarrow$ computeSCM($G$)
\STATE i $\leftarrow$ 1
\ENDIF
\IF {cost $<$ minCost}
\STATE P $\leftarrow$ G
\STATE minCost = cost
\ENDIF
\ENDWHILE
\end{small}
\end{algorithmic}
\end{algorithm}

\begin{figure}[tb!]
\centering
\vspace{-10pt}
\includegraphics[width=0.6\textwidth]{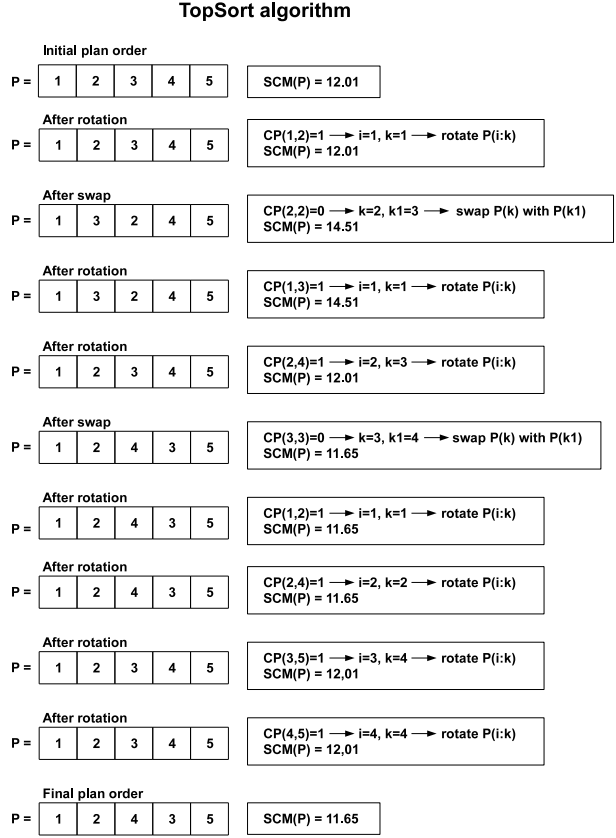}
\caption{Example of TopSort algorithm.}
\label{fig:topOrderExamp}
\end{figure}

The algorithm's pseudocode is presented in Algorithm \ref{alg:topSort}. The algorithm exhaustively checks all the permutations that satisfy the precedence constraints, and as such, it always finds the optimal solution for linear flows. The {\it computeSCM} function needs to be constructed in a way that does not compute the cost of each ordering from scratch, which is too naive, but leverages the computations of the previous plans taking into account the local changes in the new plan.
In Figure \ref{fig:topOrderExamp}, an example of finding the optimal plan of a flow using \emph{TopSort} is presented. In this example, the running steps of \emph{topSort} algorithm are depicted, given as input a valid flow execution plan (\emph{Initial plan order} plan label) and assuming the metadata of Figure \ref{fig:algInput}. Each of the given plans describe a plan generated after either  a rotation or a swap action. The optimal flow execution plan is the one labeled \emph{Final plan order}.

Note that we can implement \emph{TopSort} in a different way, where the tasks are checked from right to left. Although in \cite{VR81} this flavour is claimed to be capable of yielding better performance, this has not been verified in our flows.

\section{Extra material about the existing approximate algorithms}
\label{sec:app-appr}

\begin{algorithm}[tb!]
\caption{Swap}
\label{alg:swap}
\begin{algorithmic}[1]
\begin{small}
\REQUIRE A set of n tasks, T=\{$t_1$, ..., $t_n$\} \\
         A directed acyclic graph PC with precedence constraints
\ENSURE A directed acyclic graph P representing the optimal plan
\STATE P $\leftarrow$ randomValidPlan(PC) \COMMENT{Initiliaze P}
\STATE $swapping \leftarrow true$
\WHILE {$(swapping == true)$}
\STATE {$swapping \leftarrow false$}
\FOR {all tasks $t_i$ $\in T$}
\IF {$t_{i+1}$ has not as prerequisite $t_i$}
\IF {($computeSCM(t_i$ $\rightarrow$ $t_{i+1}$) $<$ $computeSCM(t_{i+1}$ $\rightarrow$ $t_{i}$)}
\STATE swap $t_i$ and $t_{i+1}$ in P
\STATE $swapping \leftarrow true$
\ENDIF
\ENDIF
\ENDFOR
\ENDWHILE
\end{small}
\end{algorithmic}
\end{algorithm}

\begin{figure}[tb!]
\centering
\vspace{-10pt}
\includegraphics[width=0.6\textwidth]{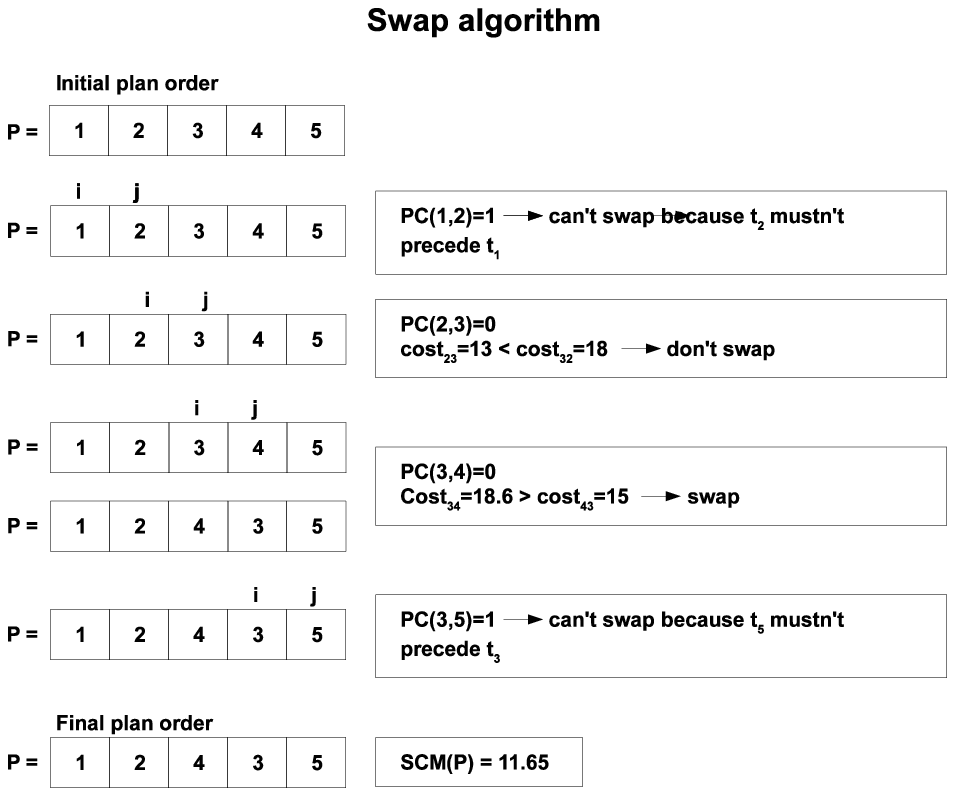}
\caption{Example of Swap algorithm.}
\label{fig:swapExampl}
\end{figure}

\begin{algorithm}[tbh!]
\caption{GreedyI}
\label{alg:chain}
\begin{algorithmic}[1]
\begin{small}
\REQUIRE A set of n tasks, T=\{$t_1$, ..., $t_n$\} \\
         A directed acyclic graph PC with precedence constraints
\ENSURE A directed acyclic graph P representing the optimal plan
\STATE P $\leftarrow \emptyset$
\STATE Cand   $\leftarrow \emptyset$  \COMMENT{Cand holds the candidate tasks}
\STATE C   $\leftarrow \emptyset$  \COMMENT{C holds the considered tasks already in P}
\STATE updateCandidates ($Cand, PC, C, T$)
\WHILE {list Cand is not empty}
\FOR {all tasks $t_j$ in Cand}
\STATE Find task $t_j$ with maximum cost where cost=(1-$sel_j$)/$cost_j$
\ENDFOR
\STATE Add $t_j$ task to optimal plan P
\STATE $C \leftarrow C \cup S_{j}$
\STATE updateCandidates ($Cand, PC, C, T$)
\ENDWHILE
\end{small}
\end{algorithmic}
\end{algorithm}

\begin{algorithm}[tbh!]
\caption{Function updateCandidates}
\label{alg:cand}
\begin{algorithmic}[1]
\begin{small}
\STATE updateCandidates ($Cand, PC, C, T$)
\FOR{all tasks $t_i$ in T}
\IF {task $t_i \not \in C$}
\IF {task $t_i$ has no prerequisites}
\STATE Add task $t_i$ to list $Cand$
\ELSE
\IF {all of the prerequisites $\in C$}
\STATE Add task $t_i$ to list $Cand$
\ENDIF
\ENDIF
\ENDIF
\ENDFOR
\end{small}
\end{algorithmic}
\end{algorithm}

\begin{figure}[tb!]
\centering
\vspace{-10pt}
\includegraphics[width=0.6\textwidth]{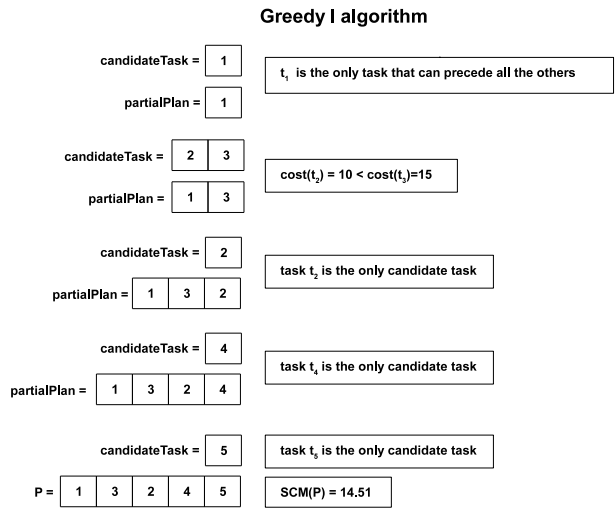}
\caption{Example of Greedy algorithm.}
\label{fig:chainExampl}
\end{figure}

\begin{algorithm}[h!]
\caption{Partition}
\label{alg:partition}
\begin{algorithmic}[1]
\begin{small}
\REQUIRE A set of n tasks, T=\{$t_1$, ..., $t_n$\} \\
         A directed acyclic graph PC with precedence constraints
\ENSURE A directed acyclic graph P representing the optimal plan
\STATE P $\leftarrow \emptyset$
\STATE Cand   $\leftarrow \emptyset$  \COMMENT{Cand holds the candidate tasks}
\STATE C   $\leftarrow \emptyset$  \COMMENT{C holds the considered tasks already in P}
\STATE updateCandidates ($Cand, PC, C, P$)
\WHILE {(Cand != $\emptyset$)}
\STATE Estimate all possible permutations of the tasks $t_i \in$ Cand
\STATE tempBestCost $\leftarrow$ 0
\STATE tempBestPlan  $\leftarrow \emptyset$
\FOR {each possible permutation $perm$}
\STATE costPerm $\leftarrow$ computeSCM($permCand$)
\IF {(costPerm $<$ tempBestCost)}
\STATE tempBestCost $\leftarrow$ costPerm
\STATE tempBestPlan  $\leftarrow $ perm
\ENDIF
\ENDFOR
\STATE Append perm to P
\STATE $C \leftarrow C \cup Cand$
\STATE updateCandidates ($Cand, PC, C, T$)
\ENDWHILE
\end{small}
\end{algorithmic}
\end{algorithm}

\begin{figure}[tb!]
\centering
\vspace{-10pt}
\includegraphics[width=0.6\textwidth]{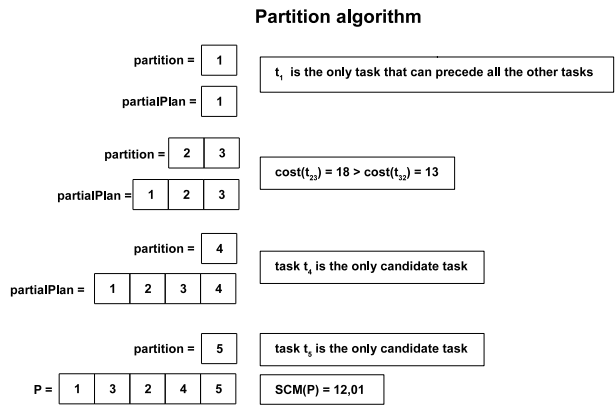}
\caption{Example of Partition algorithm.}
\label{fig:partitionExampl}
\end{figure}

Here we present the pseudocode for the Swap, GreedyI and Partition algorithms (Algorithms \ref{alg:swap}, \ref{alg:chain}, \ref{alg:partition}, respectively). Figures \ref{fig:swapExampl}, \ref{fig:chainExampl} and \ref{fig:partitionExampl} present examples for the input in Figure \ref{fig:algInput}.

\section{Extra material about the rank ordering-based techniques}
\label{sec:app-ros}

\begin{figure}[tb!]
\centering
\includegraphics[width=0.6\textwidth]{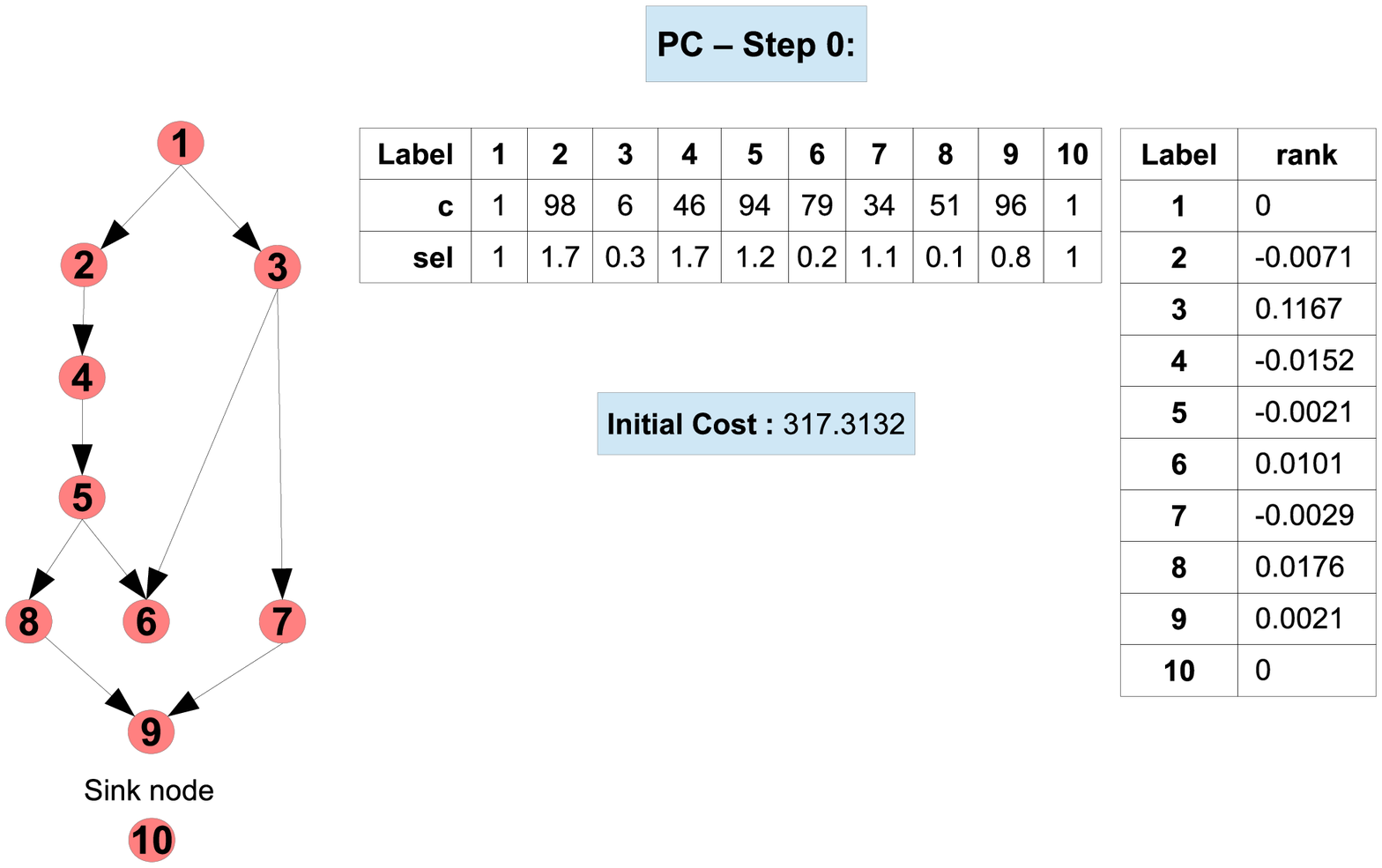}
\caption{The precedence constraint graph (PC), cost, selectivity, rank values of a data flow with 10 activities and the total execution cost.}
\label{fig:ROexampl}
\end{figure}

In this section, an illustrative example of the rank ordering methodologies is presented. Figure \ref{fig:ROexampl} depicts metadata details for a data flow with 10 tasks, which are used as input for the application of \emph{RO-I}, \emph{RO-II} and \emph{RO-III} algorithms. Specifically, this figure shows the PC graph, the values of selectivity and cost, but also the rank values that corresponds to each task of the flow. We should mention that the sink node of the data flow is disconnected from the flow in the precedence constraint graph, as it is assumed that all the flow tasks must precede this task, and we connect it after the optimization procedure is finished. The detailed examples of the rank ordering proposals are described in extend in the following.

\begin{figure}[tb!]
\centering
\includegraphics[width=0.65\textwidth]{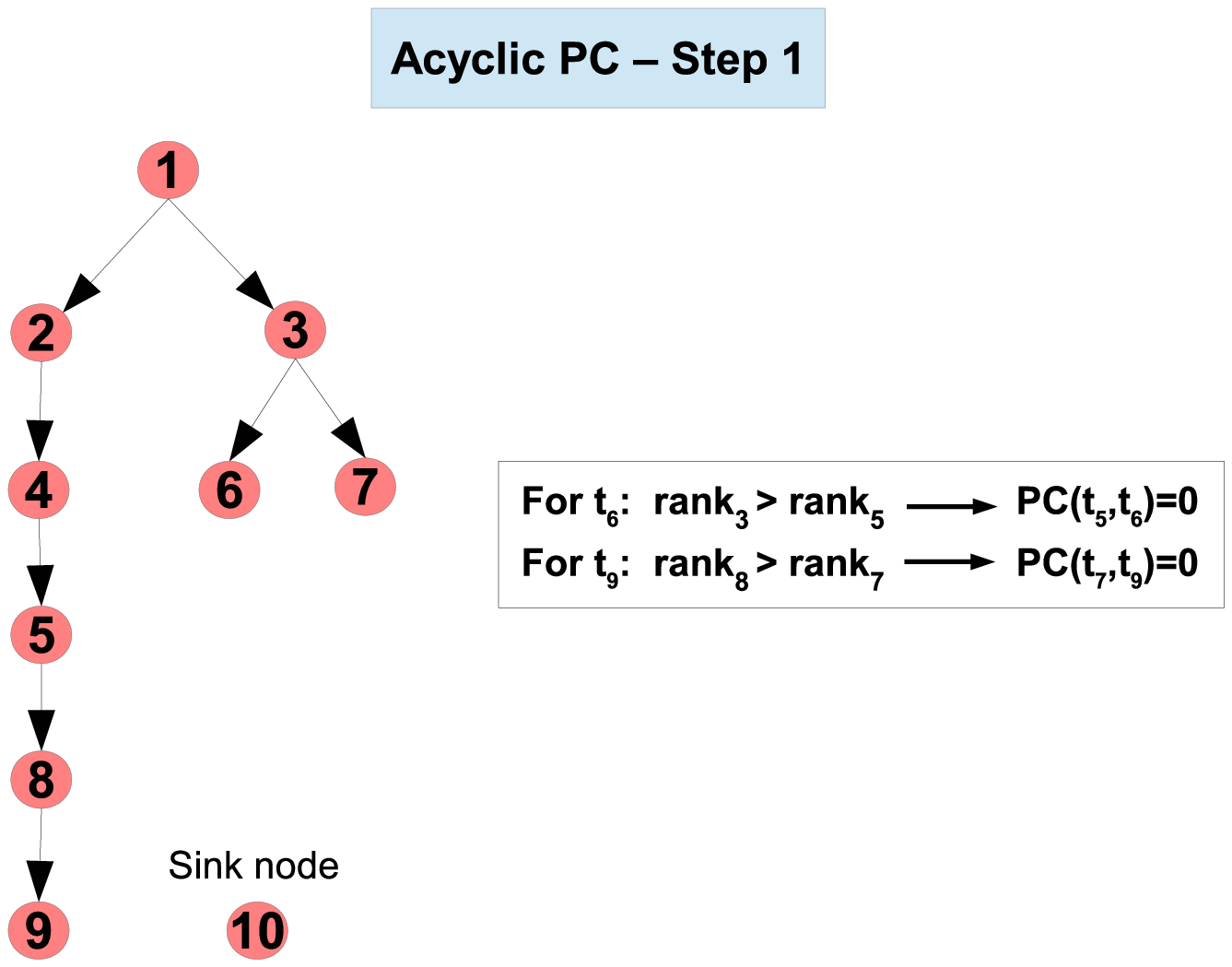}
\caption{The pre-processing phase of \emph{RO-I} to ensure that there are not cycles in the PC graph.}
\label{fig:ROIexample1}
\end{figure}

\begin{figure}[tb!]
\centering
\includegraphics[width=0.6\textwidth]{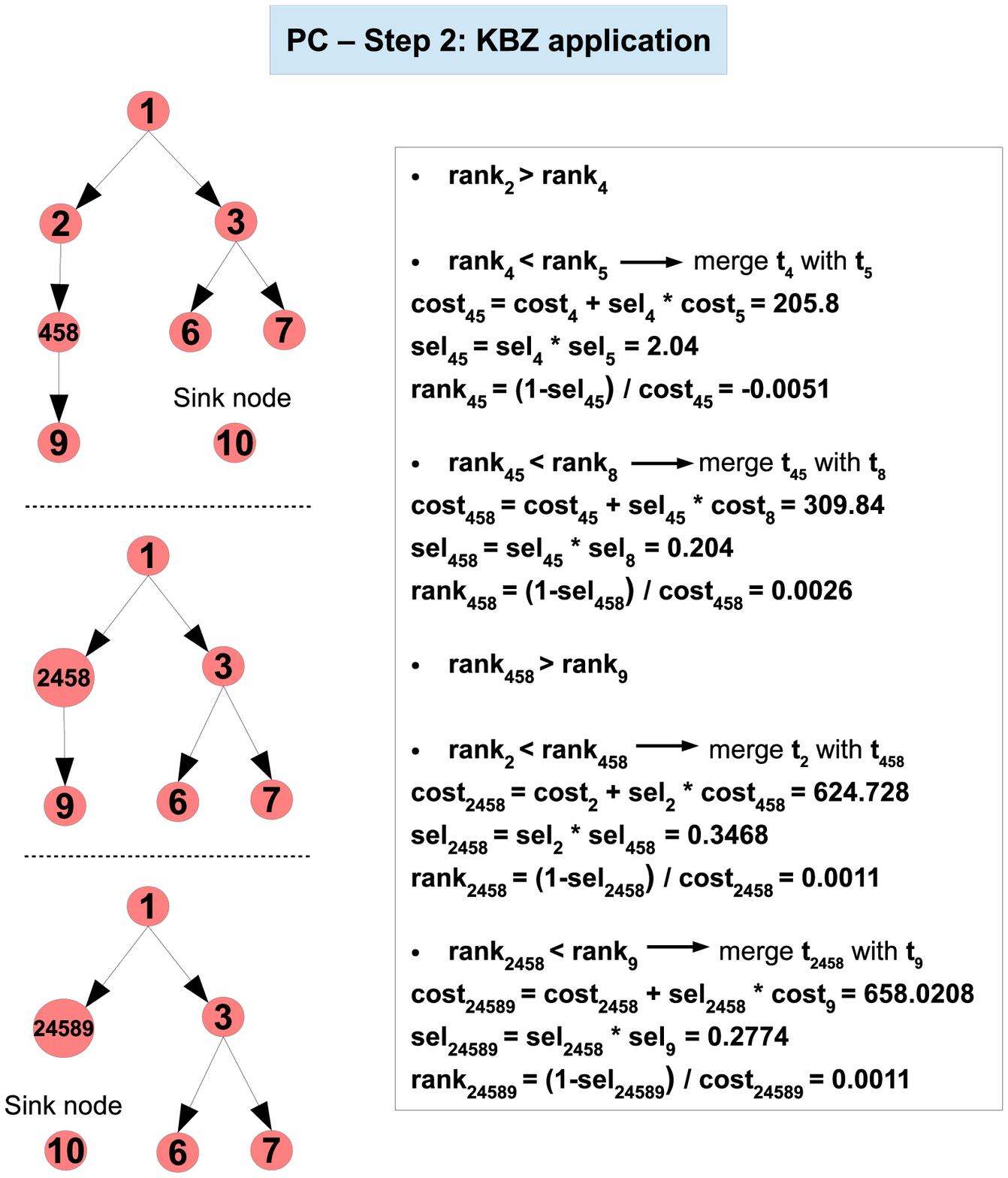}\\
\includegraphics[width=0.65\textwidth]{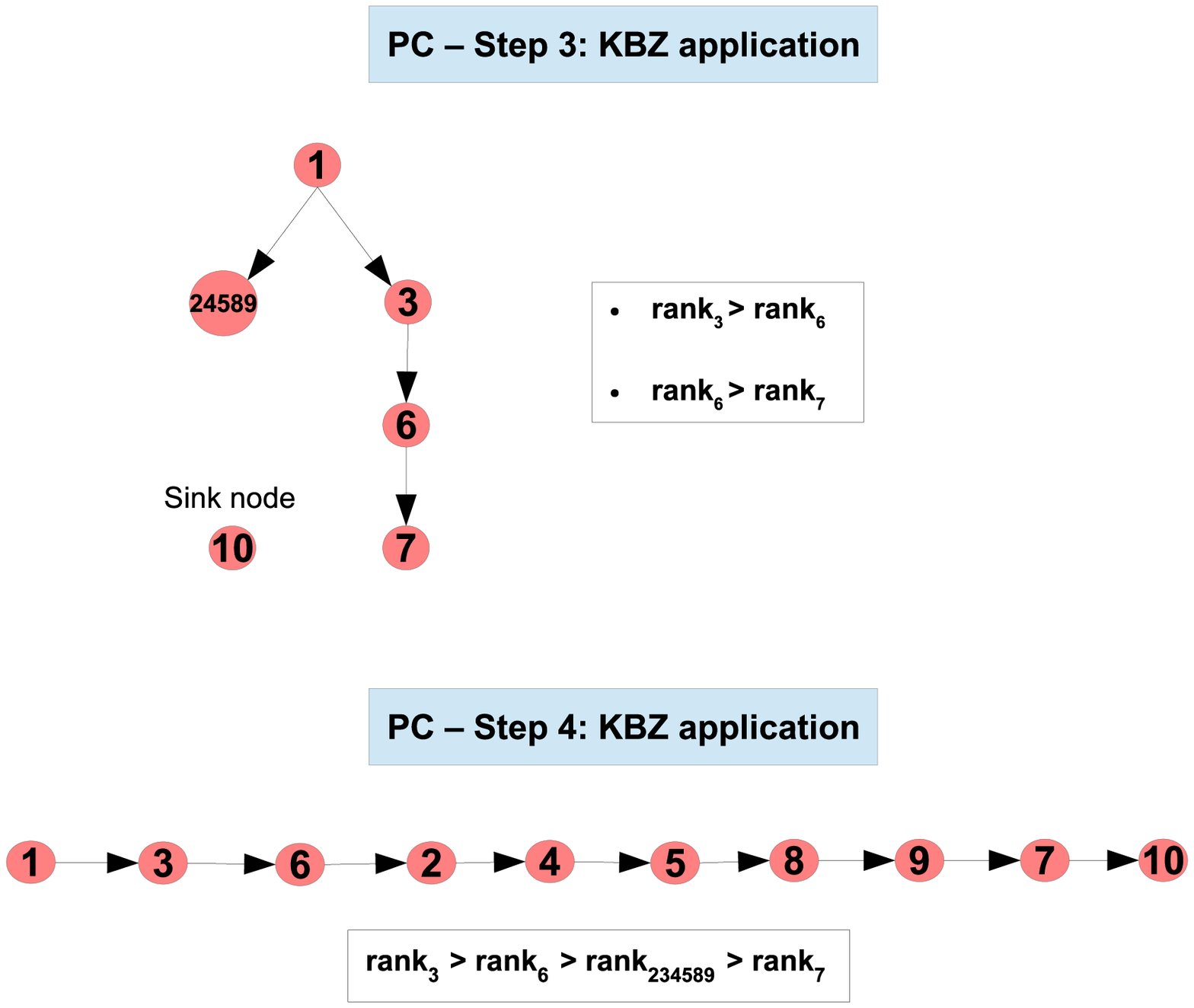}
\caption{The optimization phase of \emph{RO-I} by applying the \emph{KBZ} algorithm.}
\label{fig:ROIexample2}
\end{figure}

In Figure \ref{fig:ROIexample1}, we present the pre-processing phase of the \emph{RO-I}, in order to transform the precedence constraint graph into tree-shaped graph. The graph of the figure shows the final result of the dependency constraint graph. Then, we apply the \emph{KBZ} algorithm, which is depicted in \ref{fig:ROIexample2}.

\begin{figure}[tb!]
\centering
\includegraphics[width=0.65\textwidth]{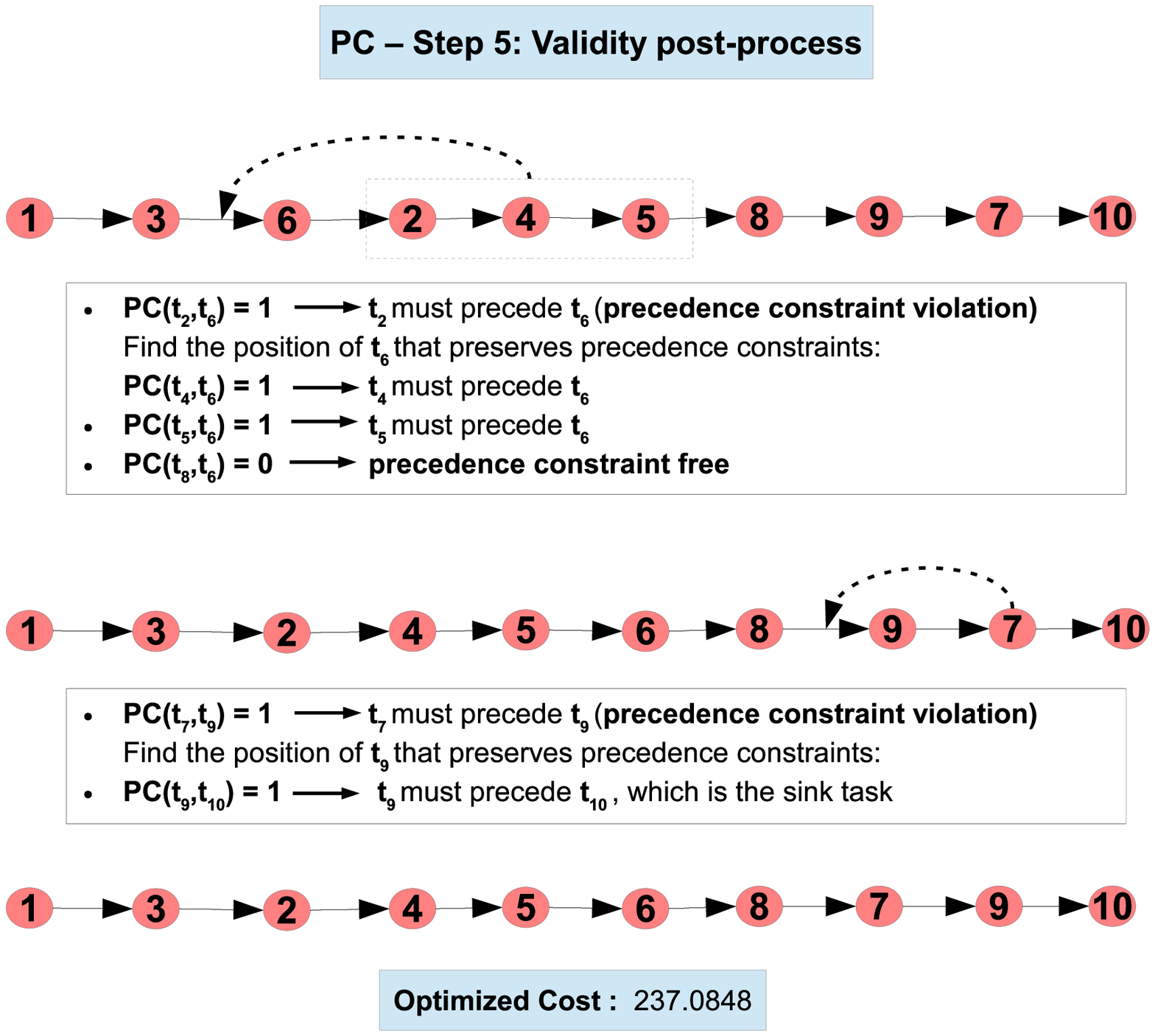}
\caption{The validity phase of \emph{RO-I} that ensures that there are not precedence constraint violations in the optimized execution plan.}
\label{fig:ROIexample5}
\end{figure}

In the following, the validity post-process phase of \emph{RO-I} is analyzed in \ref{fig:ROIexample5} and ensures that the optimized execution flow plan does not violate the dependency constraints. Finally, as is shown in this figure, the cost of the optimized execution plan is 237.0844.

\begin{figure}[tb!]
\centering
\includegraphics[width=0.8\textwidth]{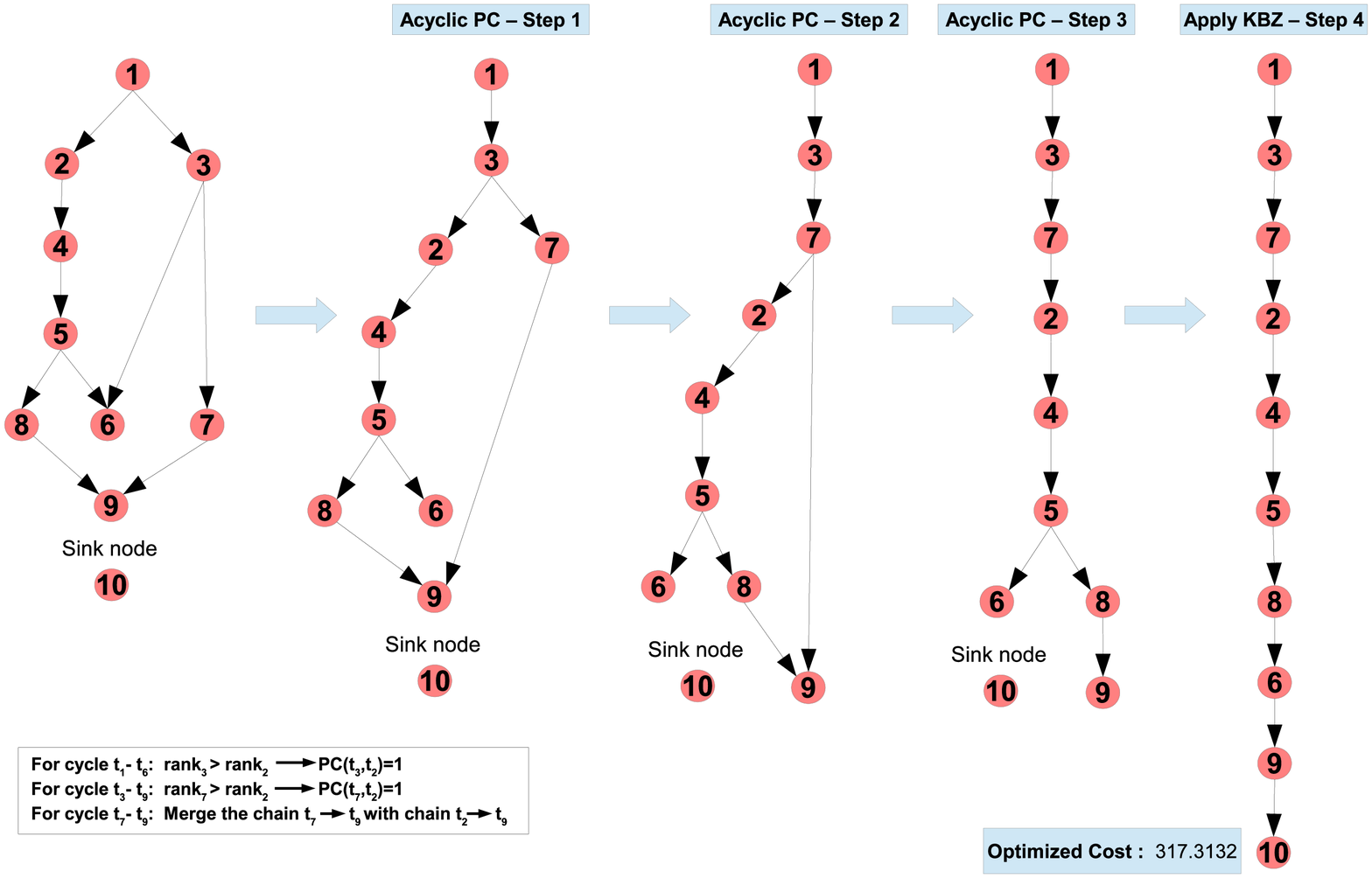}
\caption{An application example of \emph{RO-II} with the metadata of Figure \ref{fig:ROexampl}}
\label{fig:ROIIexample}
\end{figure}

The Figure \ref{fig:ROIIexample} illustrates in detail the steps of the application of \emph{RO-II}. The steps 1-3 describe the pre-processing phase of \emph{RO-II}, where we merge two sub-segments into a linear sub-flow, because they create cycles by sharing the same intermediate source and sink. The cost of the optimized flow plan returned by \emph{RO-II} methodology is 317.3132.

\begin{figure}[tb!]
\centering
\includegraphics[width=0.75\textwidth]{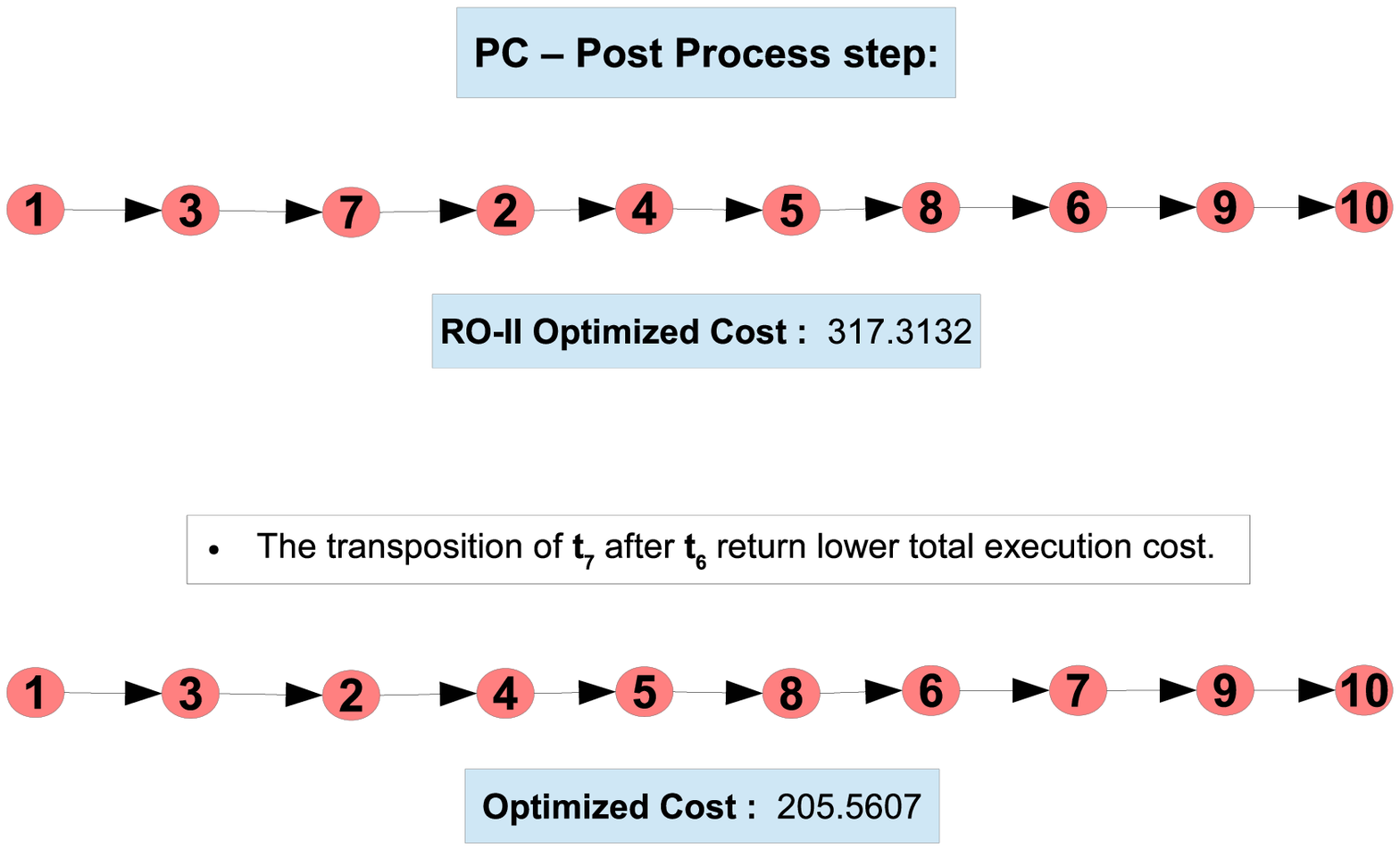}
\caption{The post-process phase of the \emph{RO-III} algorithm taking as input the generated optimized execution plan of \emph{RO-II}, as depicted in \ref{fig:ROIIexample}.}
\label{fig:ROIIIexample1}
\end{figure}

In Figure \ref{fig:ROIIIexample1}, the result of the post-processing phase of algorithm \emph{RO-III} is described. In this phase the optimized flow plan occurred by moving the flow task $t_7$ to a later stage. The optimized cost of the flow execution is 205.5607.

\newpage
\section{Extra material about the PGreedy algorithms}
\label{sec:app-pgreedy}

\begin{algorithm}[tbh!]
\caption{PGreedy}
\begin{algorithmic}[1]
\begin{small}
\REQUIRE A set of n tasks, T=\{$t_1$, ..., $t_n$\} \\
         A directed acyclic graph PC with precedence constraints
\ENSURE A directed acyclic graph P representing the optimal plan
\STATE Initialize an adjacency matrix P of optimal plan as empty
\STATE Initialize a list Cand of candidate tasks as empty
\STATE Initialize a list C of considered tasks as empty
\STATE updateCandidates ($Cand, PC, C, P$)
\WHILE {list Cand is not empty}
\FOR {all tasks $t_j$ in Cand}
\STATE $v_j$ $\leftarrow$ optimal value using a linear programming technique, which determines the optimal cost of adding $t_j$ in optimal plan P
\STATE $Cut_j$ $\leftarrow$ optimal cut for adding $t_j$ \COMMENT {\textbf{cut}: set of tasks that are the immediate predecessors}
\ENDFOR
\STATE $t_{opt}$ $\leftarrow$ task having the least $v_j$
\STATE $Cut_{opt}$ $\leftarrow$ optimal cut for adding $t_{opt}$
\STATE Add $t_{opt}$ task to optimal plan P while directed edges from the tasks in $Cut_{opt}$ to $t_{opt}$
\STATE $C \leftarrow C \cup T_{opt}$
\STATE updateCandidates ($Cand, PC, C, P$)
\ENDWHILE
\STATE computeCost($P, costs, selectivities$)
\end{small}
\end{algorithmic}
\label{alg:PGreedyI}
\end{algorithm}

\begin{figure}[tb!]
\centering
\vspace{-10pt}
\includegraphics[width=0.6\textwidth]{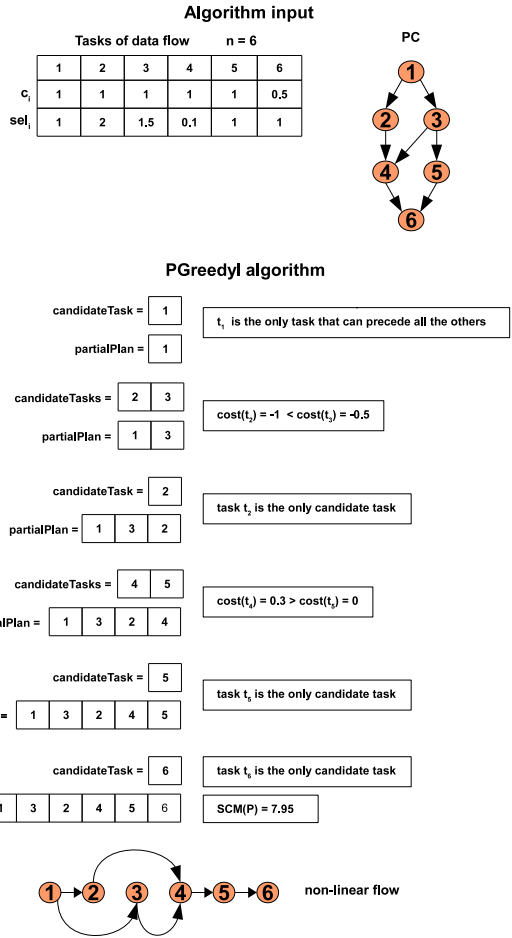}
\caption{Example of PGreedy algorithm.}
\label{fig:pGreedyExampl}
\vspace{-0.5cm}
\end{figure}

\emph{PGreedyI} algorithm is shown in Algorithm \ref{alg:PGreedyI}. In this methodology the computation of each task cost was considered by two flavors. The first one is similar with the cost metric in \cite{Sriv06}, where the cost of the task is defined as equal to $inp_i c_i$ in each step. In this case, we add the candidate task that minimizes the $inp_i c_i$ to the optimal partial plan . In the second flavour \emph{PGreedyII} the cost metric becomes $(1-sel_{i})/(inp_i c_i)$. This metric takes into account the selectivity of the next service to be appended in the execution plan and not only the selectivity of the preceding services. In Figure \ref{fig:pGreedyExampl}, an example of the \emph{PGreedyI} algorithm application based on the second cost metric is analyzed, given the cost, selectivity values, but also the precedence constraints.

\begin{figure}[tb!]
\centering
\begin{minipage}{0.4\textwidth}
\includegraphics[width=1.1\textwidth]{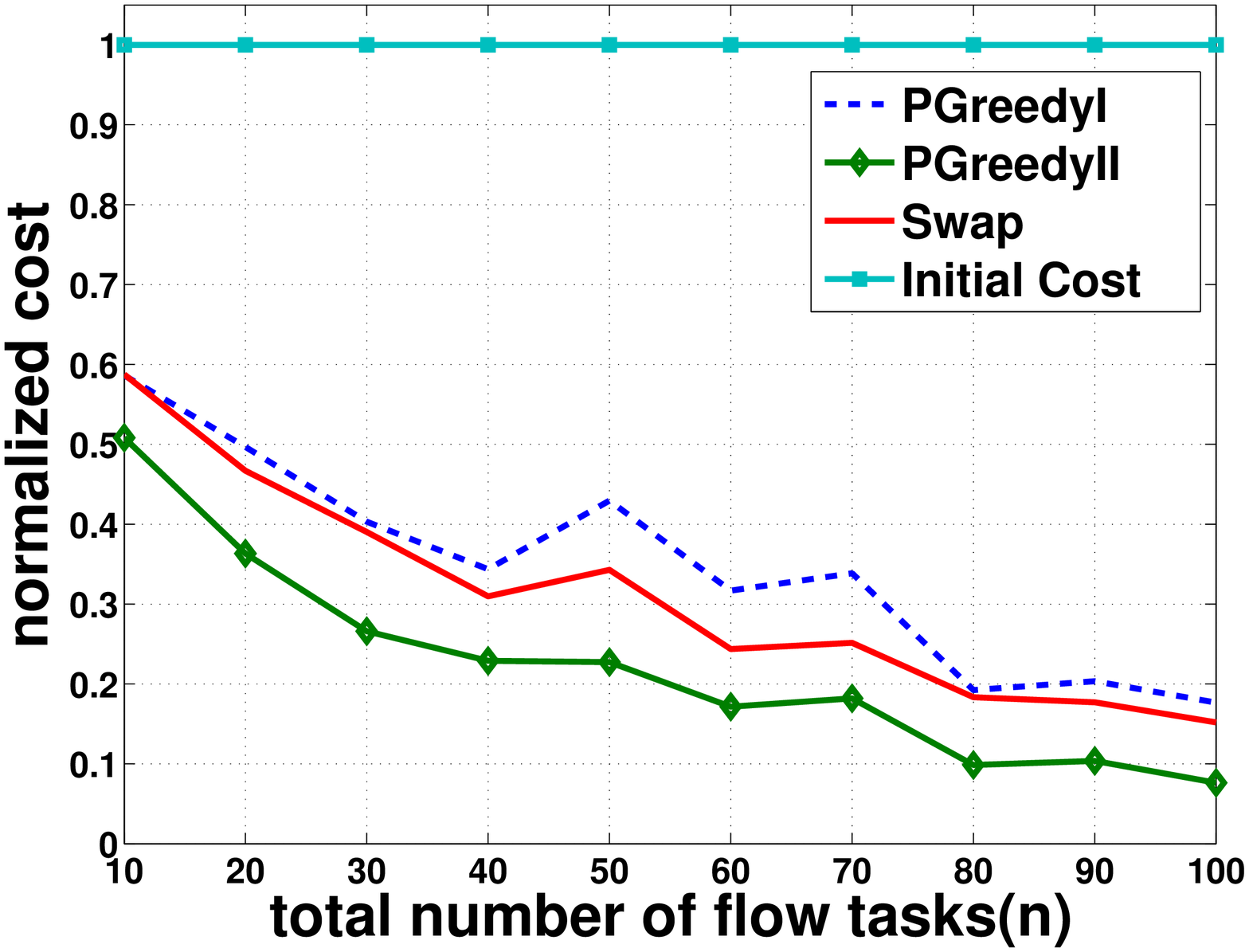}
\end{minipage}
\begin{minipage}{0.4\textwidth}
\includegraphics[width=1.1\textwidth]{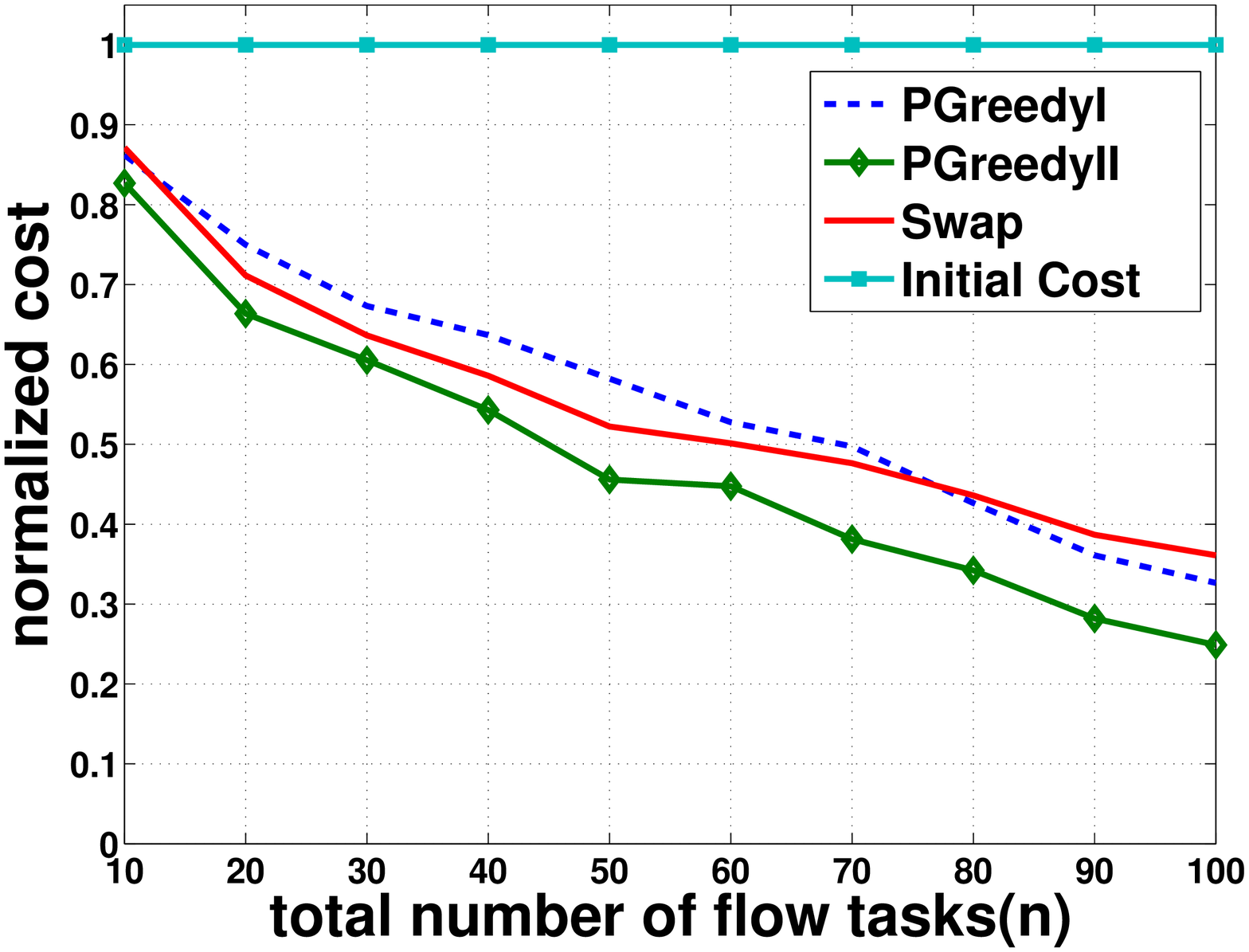}
\end{minipage}
\caption{Performance improvement for data flows with $n \in [10,100]$ with 40\% precedence constraints (left) and 80\% precedence constraints (right).}
\label{fig:parallelFlavours}
\vspace{-0.3cm}
\end{figure}

In Figure \ref{fig:parallelFlavours}, the evaluation results of the performance improvement of the \emph{PGreedy} flavours are shown. In this experiment, we compare our proposal of \emph{PGreedy} optimization algorithm with its rank-based flavour, denoted as \emph{PGreedyII}, but also each of these flavours is compared with the \emph{Swap} heuristic and the initial plan cost. The presented performance results of Figure \ref{fig:parallelFlavours} are normalized by the cost of the initial randomly generated flow execution plan. In Figure \ref{fig:parallelFlavours}(left), the \emph{PGreedy} has up to 95\% better performance improvement than the initial plan cost, whereas the execution cost of \emph{PGreedyII} can be up to 97\% lower than the initial one. In most of the iterations, \emph{PGreedyRank} seems to be clear winner. In the worst case, \emph{PGreedyII}improves the performance of the non-optimized plan by no less than 54\% on average. Also, \emph{Swap} in the best case has up to 89\% better performance
improvement than the initial flow plan. For 80\% precedence constraints, as Figure \ref{fig:parallelFlavours} shows, the \emph{PGreedyII} algorithm outperforms the other algorithms in all the data flows scenarios, even if the performance improvement decreases on average because of the limited possible reorderings. Specifically, in the best case, which is a flow with 70 tasks, \emph{PGreedyRank} has 74\% lower execution cost, while \emph{Swap} improves the initial execution cost by 58\%.

\end{document}